\documentclass{jfm}

\usepackage{graphicx}
\usepackage{newtxtext}
\usepackage{newtxmath}
\usepackage{natbib}
\usepackage{hyperref}
\hypersetup{
    colorlinks = true,
    urlcolor   = blue,
    citecolor  = black,
}

\newcommand{\RomanNumeralCaps}[1]
\linenumbers

\usepackage{pdflscape}
\usepackage{xcolor}
\usepackage{tikz}
\usepackage{array}
\usepackage{multirow}
\usepackage{capt-of}
\usepackage{afterpage}

\newcommand{\PreserveBackslash}[1]{\let\temp=\\#1\let\\=\temp}
\newcolumntype{C}[1]{>{\PreserveBackslash\centering}p{#1}}
\newcolumntype{R}[1]{>{\PreserveBackslash\raggedleft}p{#1}}
\newcolumntype{L}[1]{>{\PreserveBackslash\raggedright}p{#1}}

\definecolor{gray}{rgb}{0.5,0.5,0.5}
\newcommand{\solidline}[1][black]{\raisebox{2pt}{\tikz{\draw[-,color=#1,solid,line width = 1pt](0,0) -- (6mm,0);}}} 
\newcommand{\dashedline}[1][black]{\raisebox{2pt}{\tikz{\draw[-,color=#1,dashed,line width = 1pt](0,0) -- (6mm,0);}}}
\newcommand{\dashdotline}[1][black]{\raisebox{2pt}{\tikz{\draw[-,color=#1,dash dot,line width = 1pt](0,0) -- (6mm,0);}}}

\makeatletter
\usetikzlibrary{calc}
\newcommand*\circled[1][]{\tikz[baseline=(char.base)]{
		\node[shape=circle,draw=black,fill=white,inner sep=0pt,line width=1pt,minimum height={\f@size*1},] (char) {\vphantom{WAH1g}{\footnotesize{\textbf{#1}}}};}}
\makeatother

\graphicspath{{./Figures/}}

\title{Characterisation of rough-wall drag in compressible turbulent boundary layers}

\author{D. D. Wangsawijaya\aff{1}
\corresp{\email{D.D.Wangsawijaya@soton.ac.uk}}, 
R. Baidya\aff{2,3}, 
S. Scharnowski\aff{2}, 
B. Ganapathisubramani\aff{1}, 
\and C. J. K{\"a}hler\aff{2}}

\affiliation{
\aff{1}University of Southampton, University Road SO17 1BJ, United Kingdom
\aff{2}University of the Bundeswehr Munich, Werner-Heisenberg-Weg 39, 85577 Neubiberg, Germany
\aff{3}University of Melbourne, Grattan Street, Parkville VIC 3010, Australia
}

\begin{document}
\maketitle

\begin{abstract}

In compressible turbulent boundary layers (TBLs), roughness drag is typically characterised by first applying a velocity transformation to account for compressibility, after which the momentum deficit $\Delta U^+$ \citep{hama1954} and the equivalent sand-grain roughness $k_s$ are inferred. In practice, $k_s$ is often obtained from measurements at a single Mach number $M$ and Reynolds number $Re$, effectively forcing the roughness into the $\Delta U^+$–$\log(k_s)$ relation of \cite{nikuradse1933}. This raises a key question: \emph{if a rough surface has a known $k_s$ in incompressible flow, under what conditions can this value be used in compressible flows}? This question is explored using data obtained through a series of experiments of TBLs on rough walls (P60- and P24-grit sandpapers) over  $0.3 \leq M \leq 2.9$ and $7427 \leq Re_{\tau} \leq 30292$, including independent variation of $Re_{\tau}$ at $M=2$. Results show that $\Delta U^+$ is largely insensitive to the velocity transformation, but the fully rough regime exhibits a Mach-number-dependent shift in the logarithmic relation. Three empirical scalings are examined: an equivalent incompressible $k_s$, a viscosity-scaled roughness $k_* = k/\nu_\infty^+$ with $\nu_\infty^+ = \nu_\infty/\nu_w$, and a correction factor $\sqrt{1/F_c}$ where $F_c$ depends on $T_\infty/T_w$. The last provides the most consistent improvement across datasets, although all corrections remain empirical and rely on smooth-wall compressibility transformations. This paves the way for future work to develop custom transformation for a rough-wall TBL that can account for roughness properties and other parameters including wall conditions.

\end{abstract}



\section{Introduction}
\label{sub:intro}

A turbulent boundary layers (TBL) developing over a rough wall is a phenomenon that may occur in both incompressible and compressible flow regimes. In the incompressible flow regime, this comprises of phenomena in the transport industry, such as biofouling on ship hulls and icing on aircraft wings, as well as geophysical flows (atmospheric boundary layers developing over natural, urban landscapes, and water waves), to name a few examples. In the compressible flow regime, roughness (including steps and gaps) is likely to be induced by ablation, dust impact, water, ice droplets, and thermal expansion during the operation of high-speed flight vehicles.

\subsection{Rough walls in incompressible flows}

In the incompressible flow regime, drag penalty due to surface roughness is relatively well-defined. It is characterised by a log-law deficit from that of a smooth surface, corresponding to the \cite{hama1954} roughness function $\Delta U^+$ 
\begin{equation}
    U^+ = \frac{1}{\kappa} \log{(y+d)^+} + B - \Delta U^+
    \label{eq:rough}
\end{equation}
where $\kappa$ is the von K{\'a}rm{\'a}n constant, $U^+ = U/U_{\tau}$ is the viscous-scaled mean streamwise velocity, $U_{\tau} = \sqrt{\tau_w/\rho}$ is the friction velocity, $y^+ = y U_{\tau}/\nu$ is the viscous-scaled wall-normal coordinates, $d$ is the zero-plane displacement, $\nu$ is the kinematic viscosity, and $B$ is the log-law intercept. A constant $\kappa = 0.39$ and the log-law intercept $B = 4.3$, similar to those of \cite{squire2016} and \cite{gul2021}, are applied to all test cases in this study. 

In the so-called ``fully rough" regime, where the major contribution to the total drag comes from that of the pressure drag of the test surface's roughness elements, $\Delta U^+$ is solely a function of the equivalent sand-grain roughness $k_s$ \citep{nikuradse1933} 
\begin{equation}
    \Delta U^+ = \frac{1}{\kappa} \log{k_s^+} + B - B_{FR}
    \label{eq:hama}
\end{equation}
where $k_s^+ = k_s U_{\tau}/\nu$ and $B_{FR} = 8.5$ is the fully rough intercept. It should be noted that $k_s$ is not a physical description of a roughness. Rather, it is a measure of the effect of a roughness to the flow, relative to a sand grain-type roughness. For a matched $Re_{\tau} = \delta U_{\tau}/\nu$ ($\delta$ is the boundary-layer thickness) between flows developing over a smooth and a rough walls, $\Delta U^+$ is related to the skin friction coefficient $C_f = 2 (U_{\tau}/U_{\infty})^2$ by
\begin{equation}
    \Delta U^+ = \left. \sqrt{\frac{2}{C_f}} \right|_{smooth} - \left. \sqrt{\frac{2}{C_f}} \right|_{rough}
    \label{eq:hama_cf}
\end{equation}

\begin{figure}
    \centering
    \includegraphics[width=13.5cm, keepaspectratio]{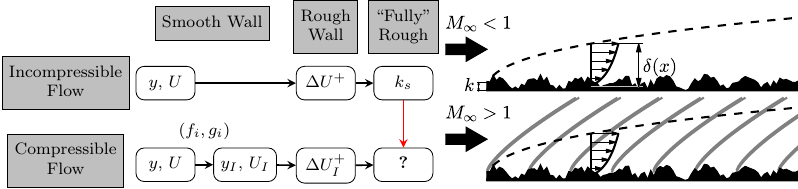}
    \caption{Established framework for drag characterisation of rough-wall turbulent boundary layers in both incompressible and compressible flow regimes.}
    \label{fig:framework}
\end{figure}

The current framework for the drag characterisation of incompressible, rough-wall TBLs, illustrated in figure \ref{fig:framework}, is well-established: from the rough-wall $U^+$ obtained by flow measurements or simulations (as a function of $y^+$), $\Delta U^+$ is calculated as a vertical shift from the smooth-wall logarithmic profile $1/\kappa \log{(y+d)^+} + B$ (equation \ref{eq:rough}). If the flow falls into the fully rough regime, $\Delta U^+$ is solely a function of $k_s$, which can be obtained directly from equation \eqref{eq:hama}. 

\begin{table}
     \begin{center}
\def~{\hphantom{0}}
    \begin{tabular}{c c c c}
    \hline
    Transformation &  Abbreviation & $f_I$ & $g_I$\\
    \hline
    \cite{howarth1948} & `HO' & \(\displaystyle \sqrt{\frac{\rho^+}{\mu^+}}  \) & 1 \\
    \cite{vandriest1951} & `VD' & 1 & \(\displaystyle \sqrt{\rho^+} \) \\ 
    \cite{brun2008} & `BR' & \(\displaystyle \frac{1}{\mu^+} \) & \(\displaystyle \sqrt{\frac{\rho^+}{\mu^+}} \frac{y}{y_I} \) \\ 
    \cite{trettel2016} & `TL' & \(\displaystyle \frac{\mathrm{d}}{\mathrm{d}y} \left( \frac{y \sqrt{\rho^+}}{\mu^+} \right) \) &  \(\displaystyle \mu^+ \frac{\mathrm{d}}{\mathrm{d}y} \left( \frac{y \sqrt{\rho^+}}{\mu^+} \right) \) \\
    \cite{volpiani2020} & `VO' & \(\displaystyle \sqrt{\frac{\rho^+}{\mu^{+^{3}}}} \) & \(\displaystyle \sqrt{\frac{\rho^+}{\mu^+}}  \) \\
    \hline
    \end{tabular}
    \caption{Various transformation functions $f_I$ and $g_I$ related to equation (\ref{eq:transform}), $\rho^+ = \rho/\rho_w$ and $\mu^+ = \mu/\mu_w$ are the density and dynamic viscosity profiles as a function of wall-normal coordinates, relative to the density and dynamic viscosity at the wall.}
    \label{tab:transform}
    \end{center}
\end{table}

\subsection{Rough walls in compressible flows}

For a wall-bounded turbulent flow developing over a \emph{smooth} surface in the compressible flow regime, the mean velocity profile $U$ and wall-normal coordinates $y$ of the TBL must be transformed or ``stretched" to account for the compressibility effect, such that the stretched profile $U_I$ and coordinates $y_I$ collapse onto the log-law of a TBL in the incompressible flow regime. In \cite{volpiani2020} and \cite{modesti2022}, generic transformations functions $f_I$ and $g_I$ transform $y$ and $U$ into their incompressible equivalents 
\begin{align}
y_I &= \int_{0}^{\hat{y}} f_I(\hat{y}) \; \mathrm{d}\hat{y} \quad \mathrm{and} \quad 
U_I = \int_{0}^{\hat{U}} g_I(\hat{U}) \; \mathrm{d}\hat{U} 
\label{eq:transform}
\end{align}
While a number of studies have been devoted to formulate various $f_I$ and $g_I$, in this study we limit the analysis and discussion for 5 different transformations listed in table \ref{tab:transform}, identical to those chosen by \cite{volpiani2020}, and some by \cite{modesti2022}. These transformations are: the Van Driest I transformation \citep[abbreviated to `VD' in this study]{vandriest1951}, which is the most widely known and used transformation, an older transformation \citep[or `HO']{howarth1948}, and 3 newer ones: \cite{brun2008} -- `BR', \cite{trettel2016} -- `TL', and \cite{volpiani2020} -- `VO'. As shown in table \ref{tab:transform}, the transformation functions $f_I$ and $g_I$ typically account for density and viscosity variations across the boundary-layer, $\rho^+ = \rho/\rho_w$ and $\mu^+ = \mu/\mu_w$, where $\rho = \rho(y)$ and $\mu = \mu(y)$ are the flow density and dynamic viscosity as functions of $y$, and the subscript `$w$' denotes quantities at the wall.  

For a rough surface exposed to a compressible inlet condition (figure \ref{fig:framework}, $M_{\infty} > 1$ is the freestream Mach number), shock waves form as the flow encounters protrusions of the roughness elements above the sonic line. Thus, in a addition to the roughness element drag in the incompressible flow regime, compressible, rough-wall TBLs generate the so-called wave drag, as illustrated in figure \ref{fig:framework}. At present, there are no $f_I$ and $g_I$ transformation functions that account for roughness-generated shock waves above a particular rough wall. Instead, $U$ and $y$ are transformed first into their incompressible equivalents, $U_I$ and $y_I$ via equation \eqref{eq:transform} and the transformation functions shown in table \ref{tab:transform}. Then, the Hama roughness function is calculated from the stretched profile $U_I$
\begin{equation}
    U_I^+ = \frac{1}{\kappa} \log (y + d)_I^+ + B - \Delta U_I^+
    \label{eq:rough_compressible}
\end{equation}
where $U_I^+ = U_I/U_{\tau}$, $(y+d)_I^+ = (y+d)_I U_{\tau}/\nu_w$, $U_{\tau} = \sqrt{\tau_w/\rho_w}$, and $\Delta U_I^+$ is the Hama roughness function obtained from the transformed velocity and wall-normal coordinates. Finally, in the case of fully rough flow, the equivalent sand grain roughness $k_s$ for a particular roughness is derived directly from its $\Delta U_I^+$, similar to equation \ref{fig:hama}, but with $\Delta U_I^+$ instead of $\Delta U^+$.

\begin{figure}
    \centering
    \includegraphics[width=12.25cm, keepaspectratio]{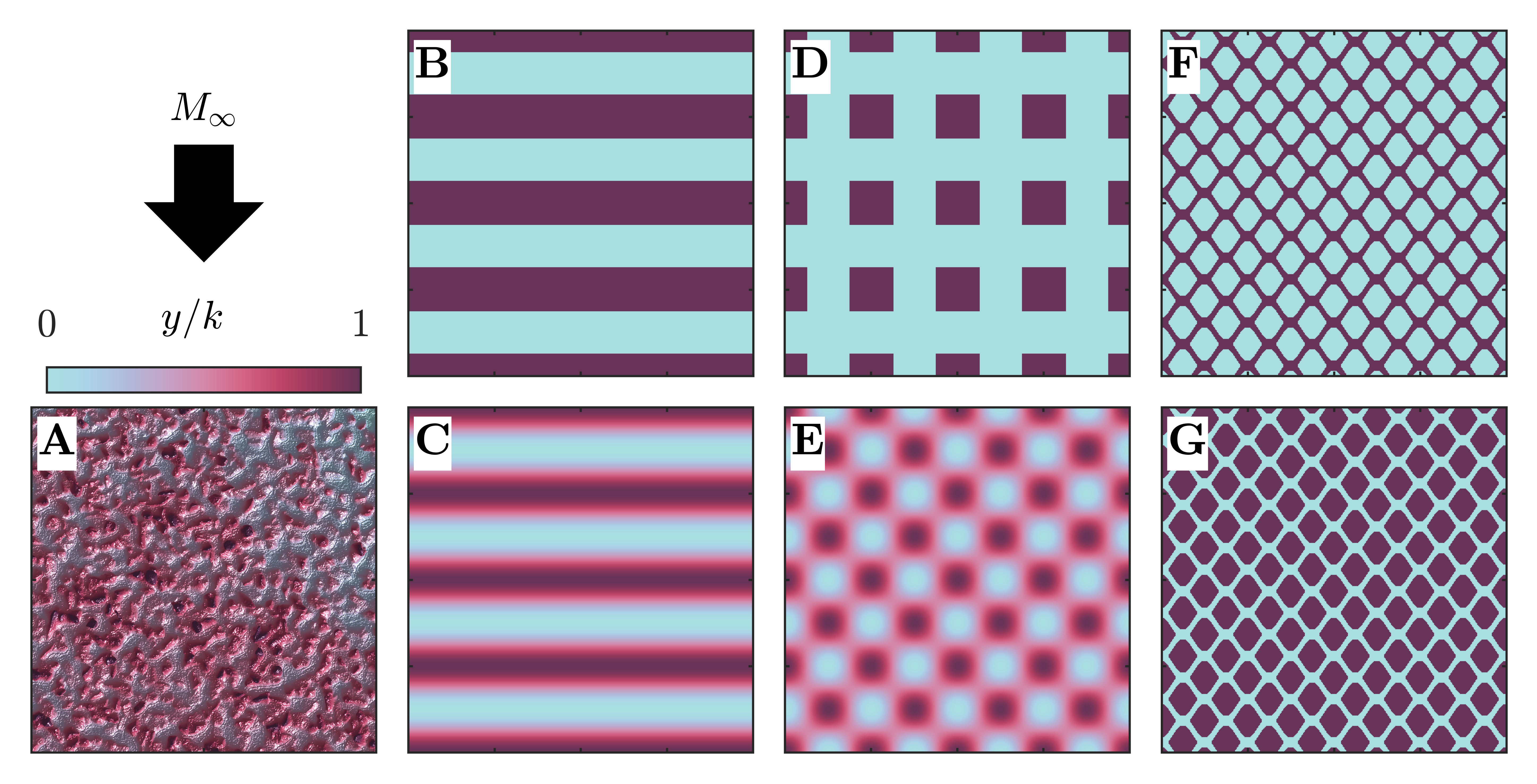}
    \caption{Coloured contours of various roughness topologies relative to the roughness element height $k$. Types of roughnesses from existing studies listed in table \ref{tab:ref}: type `\textbf{A}' (sandpaper), `\textbf{B}' (2D traverse bars), `\textbf{C}' (2D sine waves), `\textbf{D}' (cubes), `\textbf{E}' (egg carton), `\textbf{F}' (expanded mesh), and `\textbf{G}' (diamonds). Black arrow indicates the direction of the flow.}
    \label{fig:roughness_type}
\end{figure}

The present drag characerisation framework for compressible, rough-wall TBLs, summarised in figure \ref{fig:framework}, is evident in the previous studies conducted on various rough surfaces at the compressible regime. The number of available studies are limited, however, we are able to categorise these studies into 7 different roughness ``families", as illustrated in figure \ref{fig:roughness_type}: type \textbf{A} (sandpaper roughness), \textbf{B} (2D traverse bars), \textbf{C} (2D sine waves), \textbf{D} (cubes), \textbf{E} (egg carton roughness), \textbf{F} (expanded mesh), and \textbf{G} (diamonds). 

\begin{table}
     \begin{center}
    \def~{\hphantom{0}}
    \begin{tabular}{c l c c c c c c c}
    \hline
    Type & Reference & $M_{\infty}$, ($M_{b}$) & $Re_{\tau}$ & $\delta/k$ & $\Delta U_I^+$ & Wall & Drag & Sym. \\
    \hline
     \multirow{11}{*}{\textbf{A}} & \cite{goddard1959} & 2.75 & -- & 59.3 & 2.40 & \circled[c] & \circled[1] & \includegraphics[width=3mm, keepaspectratio]{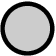}\\
     & \cite{goddard1959} & 2.75 & -- & 32.7 & 5.27 & \circled[c] & \circled[1] & \includegraphics[width=3mm, keepaspectratio]{set3/mark_circ1}\\
     & \cite{goddard1959} & 2.75 & -- & 15.3 & 7.95 & \circled[c] & \circled[1] & \includegraphics[width=3mm, keepaspectratio]{set3/mark_circ1}\\
     & \cite{reda1975}: P80 & 2.75 & -- & 173 & 1.20 & \circled[b] & \circled[1] & \includegraphics[width=3mm, keepaspectratio]{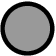}\\
     & \cite{reda1975}: P50 & 2.90 & -- & 92.6 & 4.50 & \circled[b] & \circled[1] &  \includegraphics[width=3mm, keepaspectratio]{set3/mark_circ2}\\
     & \cite{reda1975}: P36 & 2.90 & -- & 66.1 & 5.83 & \circled[b] & \circled[1] &  \includegraphics[width=3mm, keepaspectratio]{set3/mark_circ2}\\
     & \cite{reda1975}: P24 & 2.90 & -- & 40.2 &  7.67 & \circled[b] & \circled[1] &  \includegraphics[width=3mm, keepaspectratio]{set3/mark_circ2}\\
     & \cite{reda1975}: P24 & 2.90 & -- & 43.1 &  9.23 & \circled[b] & \circled[1] &  \includegraphics[width=3mm, keepaspectratio]{set3/mark_circ2}\\
     & \cite{latin2000}: P80 & 2.73 & 3475 & 27.7 & 8.33  & \circled[c] & \circled[2] &  \includegraphics[width=3mm, keepaspectratio]{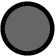}\\
     & \cite{latin2000}: P36 & 2.72 & 5007 & 20.0 & 11.58 & \circled[c] & \circled[2] & \includegraphics[width=3mm, keepaspectratio]{set3/mark_circ3}\\
     & \cite{latin2000}: P20 & 2.70 & 5104 & 21.3 & 12.48 & \circled[c] & \circled[2] & \includegraphics[width=3mm, keepaspectratio]{set3/mark_circ3}\\
    
    \hline
     \multirow{5}{*}{\textbf{B}} & \cite{berg1979} & 6.00 & 583 & 80.8 & 2.69 & \circled[c] & \circled[1] & \includegraphics[width=3mm, keepaspectratio]{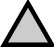}\\
     & \cite{berg1979} & 5.98 & 612 & 40.4 & 3.76 & \circled[c] & \circled[1] & \includegraphics[width=3mm, keepaspectratio]{set3/mark_tria1}\\
     & \cite{berg1979} & 5.97 & 745 & 21.6 & 6.55 & \circled[c] & \circled[1] & \includegraphics[width=3mm, keepaspectratio]{set3/mark_tria1}\\
     & \cite{latin2000} & 2.73 & 4454 & 30.0 & 10.82 & \circled[c] & \circled[2] & \includegraphics[width=3mm, keepaspectratio]{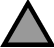}\\
     & \cite{sahoo2009} & 7.30 & 733 & 7.3 & 13.00 & \circled[a] & \circled[4] & \includegraphics[width=3mm, keepaspectratio]{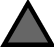}\\
     
    \hline
     \multirow{3}{*}{\textbf{C}} & \cite{tyson2013}$^{\dagger}$ & (1.5, 3) & 160, 180 & 12.5 & 4.24, 4.02 & \circled[d] & \circled[3] & \includegraphics[width=3mm, keepaspectratio]{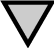}\\
     & \cite{tyson2013}$^{\dagger}$ & (1.5, 3) & 160, 180 & 6.3 & 7.09, 6.70 & \circled[d] & \circled[3] & \includegraphics[width=3mm, keepaspectratio]{set3/mark_trib1}\\
     & \cite{aghaei-jouybari2023} & (1.50) & 236, 250 & 10.0 & 6.61, 7.76 & \circled[d] & \circled[3] & \includegraphics[width=3mm, keepaspectratio]{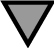}\\
     
    \hline
     \multirow{4}{*}{\textbf{D}} & \cite{latin2000} & 2.73 & 4184 & 28.2 & 10.38 & \circled[c] & \circled[2] & \includegraphics[width=3mm, keepaspectratio]{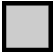}\\
     & \cite{ekoto2008} & 2.86 & 6235 & 14.7 & 12.20 & \circled[c] & \circled[4] & \includegraphics[width=3mm, keepaspectratio]{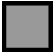}\\
     & \cite{modesti2022}$^{\dagger}$ & (2, 4) & 498--2688 & 12.5 & 5.04--8.83 & \circled[e] & \circled[3] & \includegraphics[width=3mm, keepaspectratio]{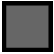}\\
     & \cite{neeb2024} & 4.9--5.2 & 307--384 & 8.6--11 & 5.62--6.34 & \circled[a] & \circled[4] & \includegraphics[width=3mm, keepaspectratio]{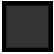}\\
     
    \hline
     \multirow{2}{*}{\textbf{E}} & \cite{aghaei-jouybari2023} & (1.50) & 225, 235 & 10.0 & 5.15, 6.30 & \circled[d] & \circled[3] & \includegraphics[width=3mm, keepaspectratio]{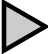}\\
     & \cite{wang2024} & 2.3--7.3 & 581--625 & 12.7--13.5 & 9.20 & \circled[d] & \circled[3] & \includegraphics[width=3mm, keepaspectratio]{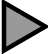}\\
     
    \hline
    \textbf{F} & \cite{sahoo2009} & 7.30 & 846 & 6.3 & 13.80 & \circled[a] & \circled[4] & \includegraphics[width=3mm, keepaspectratio]{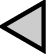}\\
    
    \hline
     \multirow{2}{*}{\textbf{G}} & \cite{ekoto2008} & 2.86 & 4575 & 13.9 & -- & \circled[c] & \circled[4] & \includegraphics[width=3mm, keepaspectratio]{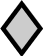}\\
     & \cite{peltier2016} & 4.89 & 2300 & 14.1 & 13.00 & \circled[a] & \circled[4] & \includegraphics[width=3mm, keepaspectratio]{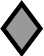}\\
    \hline
    \end{tabular}
    \caption{Summary of previous studies conducted on rough-wall in compressible, wall-bounded turbulence. Roughness types correspond to figure \ref{fig:roughness_type}. $M_{\infty}$ is the freestream Mach number, while $M_b$ (in brackets) is the bulk Mach number of channel flow studies. Column `Wall' refers to various conditions at the wall, which are \circled[a]: direct wall temperature $T_w$ measurements, \circled[b]: direct $T_w$ measurements only for a reference case, adiabatic wall was assumed for the rest of the test cases, \circled[c]: adiabatic wall was assumed, \circled[d]: isothermal wall was set as the simulation's boundary condition and \circled[e]: cooled wall was set as the simulation's boundary condition. Column `Drag' refers to various techniques to obtain the WSS, which are \circled[1]: direct WSS measurements, \circled[2]: direct WSS measurements only for a reference case, a fitting method utilised for the rest of the test cases, \circled[3]: WSS from the wall velocity gradient, and \circled[4]: WSS obtained from a fitting method. `$\dagger$' denotes available incompressible flow simulations at $M_b = 0.3$. Symbols denote various roughness types, with grey shades denoting different studies.}
    \label{tab:ref}
    \end{center}
\end{table}

The complete list of all previous studies on every type of roughness in figure \ref{fig:roughness_type} is summarised in table \ref{tab:ref}. For sandpaper roughness (figure \ref{fig:roughness_type}\textbf{A}), whenever available, the grit size is reported. A wide range of roughness element sizes ($6.3 \leq \delta/k \leq 80.8$, $k$ is the physical height of the roughness) is covered across all types of roughness shown in table \ref{tab:ref}. The majority of these studies were conducted in the supersonic flow regime, $1.5 \leq M_{\infty} < 5$ (or the bulk Mach number $M_b$ in channel flow studies), with some \citep{sahoo2009,neeb2024,wang2024} extended towards the hypersonic regime. In two studies \citep{tyson2013,modesti2022}, marked by `$\dagger$' in table \ref{tab:ref}, a baseline simulation for the same roughness was conducted in the incompressible flow regime ($M_b = 0.3$). Conditions at the wall varied: in some experimental campaigns \citep{sahoo2009,peltier2016,neeb2024}, the wall temperatures $T_w$ were measured directly with thermocouples or infrared thermography (marked by `\circled[a]' in table \ref{tab:ref}), while in another study \citep{reda1975}, direct $T_w$ measurements were only available for a reference case (typically a smooth wall), while for the rest of the roughness surfaces, adiabatic walls were assumed (`\circled[b]' in table \ref{tab:ref}). In other studies \citep{goddard1959,berg1979,latin2000,ekoto2008}, no direct measurements of $T_w$ were available and thus adiabatic walls were assumed (`\circled[c]' in table \ref{tab:ref}). In numerical simulation studies, $T_w$ was somewhat imposed as a boundary condition: either isothermal walls \citep{tyson2013,aghaei-jouybari2023,wang2024} or walls cooled by a certain cooling rate \citep{modesti2022}. These studies are marked by `\circled[d]' and `\circled[e]' in table \ref{tab:ref}, respectively.       

In terms of the wall shear stress (WSS), direct measurements using drag balance were conducted in some experimental studies \citep{goddard1959,reda1975,berg1979}, marked by `\circled[1]' in table \ref{tab:ref}, or it was obtained from the velocity gradient on the wall in numerical simulations of \cite{tyson2013,modesti2022,aghaei-jouybari2023,wang2024} (`\circled[3]' in table \ref{tab:ref}). \cite{latin2000} limited the direct WSS measurements to a reference smooth wall case (marked by `\circled[2]' in table \ref{tab:ref}), while the rough WSS was obtained using some profile fitting techniques, similar to those described in \S\ref{sub:wss}. When direct measurements were not supported in the experimental facilities \citep{ekoto2008,sahoo2009,peltier2016,neeb2024}, fitting techniques were used for the entirety of the experimental data. These are marked by `\circled[4]' in table \ref{tab:ref}. The obtained WSS, either from fitting or direct measurements, was then combined with the stretched $U_I$ and $y_I$ via equation \eqref{eq:transform} to determine $\Delta U_I^+$ (figure \ref{fig:framework}). A large majority of studies in table \ref{tab:ref} used VD transformation (formula shown in table \ref{tab:transform}), while in \cite{modesti2022}, other transformations such as TL and VO were used in addition to the VD transformation. A custom transformation was used by \cite{wang2024}. From there, the vertical shift $\Delta U_I^+$ was determined, and the reported values are listed in table \ref{tab:ref}.     

\subsection{Aim, scope, and limitations}

As shown in table \ref{tab:ref}, all experimental studies on rough walls in compressible flows were conducted at a single $M_{\infty}$ and $Re_{\tau}$ (which was determined by $M_{\infty}$). In some simulations \citep{tyson2013,wang2024}, $M_b$ sweeps were conducted, while in \cite{modesti2022} an independent $Re_{\tau}$ sweep at a constant $M_b$ was also conducted. For the measurements conducted at a single $M_{\infty}$ and $Re_{\tau}$, $k_s$ of a certain roughness was directly derived from the obtained $\Delta U_I^+$ (figure \ref{fig:framework}), implying that this $k_s$ is \emph{forced} to collapse onto Nikuradse's formula for incompressible flows \citep{nikuradse1933}.

Considering this existing framework (figure \ref{fig:framework}) for drag characterisation of rough walls in compressible flows, and the limited amount of data available in this flow regime (table \ref{tab:ref}) compared to those within the incompressible flow regime, we aim to investigate the possibility of transferring the established framework for rough-wall drag in incompressible flows to the compressible flow regime. This is illustrated by the red arrow in figure \ref{fig:framework}: \emph{if $k_s$ of a certain roughness in incompressible flows is known, what would it take to use this $k_s$ for the same roughness in compressible flows?} We note that this approach is not new; a similar approach has been attempted by \cite{modesti2022} for direct numerical simulations of a channel flow covered with cube roughness (figure \ref{fig:roughness_type}\textbf{D}), where $k_s$ from a simulation at $M_b = 0.3$ must be corrected by a factor for drag characterisation at $M_b = 2$ and 4. Therefore, in this study, we intend to fill in the remaining gaps with experimental data in a higher range of $Re_{\tau}$ ($7427 \leq Re_{\tau} \leq 30292$) for two types of realistic, sandpaper-like surface roughness (figure \ref{fig:roughness_type}\textbf{A}). A range of nominal test section Mach numbers $M$ is covered in this study, spanning subsonic ($M = 0.3$), transonic ($M = 0.8$), and supersonic flow regimes (up to $M = 2.9$), completed with an independent variation of $Re_{\tau}$ at a constant $M = 2$, as detailed in \S\ref{sub:setup}. In \S\ref{sub:corr_factors}, the new framework (figure \ref{fig:framework}) will be tested for various compressibility transformations shown in table \ref{tab:transform}, as well as data from previous studies in table \ref{tab:ref}, and the results will be discussed further in \S\ref{sub:discuss}. We acknowledge that, while the experimental facility enables independent $Re$ and $M$ sweeps, its limitations prevent direct $T_w$ and WSS measurements on the rough walls. The estimation techniques for these quantities are detailed in \S\ref{sub:temperature} and \S\ref{sub:wss}, respectively. 

\section{Methodology}
\label{sub:setup}

\subsection{Facility and air properties}

Measurements are conducted inside the intermittent, blow-down trisonic wind tunnel (TWM) facility at the University of the Bundeswehr Munich. Compressed air is pumped into the holding tanks prior to each experimental run, released into the 1800 mm $\times$ 300 mm $\times$ 675 mm (length $\times$ width $\times$ height) test section, and finally exhausted to the atmosphere. The facility is equipped with an adjustable Laval nozzle and a diffuser, allowing a range of operating Mach numbers of $0.3 \leq  M \leq 3$. Within this operational range, the freestream turbulence intensity ranges from 1.9\% to 4.5\% (inversely proportional to $M$), as recorded by \cite{scharnowski2019}. 

\begin{figure}
    \centering
    \includegraphics[width=13.5cm, keepaspectratio]{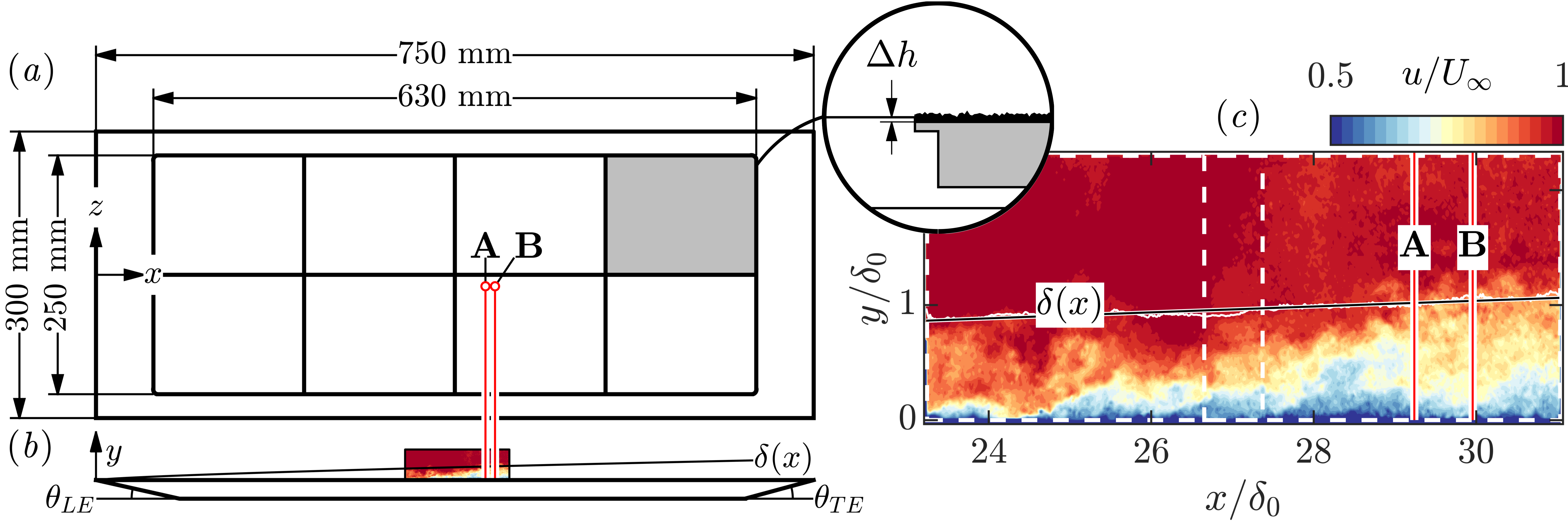}
    \caption{(a) Top view ($x$--$z$ plane) and (b) side view ($x$--$y$ plane) of the baseplate with 8 inserts where the rough surfaces are attached. Red circles and solid lines mark `\textbf{A}' and `\textbf{B}' measurement stations. Black solid lines mark the 99\% boundary-layer thickness growth $\delta(x)$, while $\theta_{LE}$ and $\theta_{TE}$ are the leading and trailing edge ramp angles relative to the $x$-direction. \textit{Inset:} cross-sectional view of the baseplate with an insert (gray-shaded), showing the insert's thickness lowered by $\Delta h$ to approximately match the height of the baseplate and the rough walls (black-shaded). (c) Coloured contours of the normalised instantaneous streamwise velocity $u/U_{\infty}$ of the reference case (SW at $M = 0.3$). White solid line shows the $\delta(x)$ obtained from PIV measurements. Black solid line marks the power law fit $0.194 x^{0.714}$ (in mm), obtained by fitting $\delta(x)$ of the reference case. White dashed lines mark the FOVs from the two PIV cameras, with 10 mm overlap in $x$-direction. $\delta_0$ is the $\delta$ averaged over $x = 407 \pm 5$ mm (station `\textbf{A}').}
    \label{fig:setup}
\end{figure}

Table \ref{tab:cases} shows key air properties of all test cases. For all test cases, the experimental data are recorded for approximately 20 s for each run. The mean stagnation pressure $P_0$ (subscript `0' denotes a total or stagnation quantity) is recorded in the settling chamber, while the mean stagnation temperature $T_0$ is measured in the holding tanks. The mean nominal test section Mach number $M$ is obtained from the isentropic relation, while $M_{\infty}$ and $U_{\infty}$ are the mean freestream Mach number and velocity obtained from PIV measurements at the measurement stations shown in figure \ref{fig:setup}(a). For the viscous-scaled quantities shown in table \ref{tab:cases}, the dynamic viscosity $\mu$ is calculated using Sutherland's formula. The complete description of the test facility is given by \cite{baidya2020}. 

\subsection{Test cases}

\begin{table}
    \begin{center}
    \def~{\hphantom{0}}
    \begin{tabular}{C{4mm} C{5mm} C{5mm} C{5mm} C{8mm} C{8mm} C{8mm} C{7mm} C{5mm} C{8mm} C{8mm} C{8mm} C{3mm} C{3mm} C{3mm} C{3mm} C{3mm}}
    \hline
    $M$ & $M_{\infty}$ & $P_0$ & $T_0$ & $T_{\infty}/T_w$ & $U_{\infty}$ & $U_{\tau_{VD}}$ & $\delta$ & $\theta$ & $Re_{\tau_{VD}}$ & $C_{f_{VD}}$ & $\Delta U_{VD}^+$ & \multicolumn{5}{c}{Symbol} \\
    $[-]$ & $[-]$ & [bar] & [K] & $[-]$ & [m/s] & [m/s] & [mm] & [mm] & $[-]$ & \scriptsize{$[\times 10^{-3}]$} & $[-]$ & HO & VD & BR & TL & VO \\
    \hline
    \multicolumn{17}{c}{\textbf{Smooth Wall (SW)}} \\
    0.3 & 0.28 & 1.5 & 293 & 0.99 & 93.82 & 3.36 & 13.92 & 1.38 & 4355 & 2.53 & -- & \includegraphics[width=3mm, keepaspectratio]{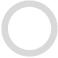} & \includegraphics[width=3mm, keepaspectratio]{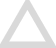} & \includegraphics[width=3mm, keepaspectratio]{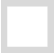} & \includegraphics[width=3mm, keepaspectratio]{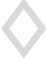} & \includegraphics[width=3mm, keepaspectratio]{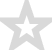} \\
    0.8 & 0.80 & 1.5 & 286 & 0.90 & 255.45 & 8.76 & 22.29 & 1.86 & 13361 & 2.11 & -- & \includegraphics[width=3mm, keepaspectratio]{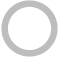} & \includegraphics[width=3mm, keepaspectratio]{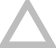} & \includegraphics[width=3mm, keepaspectratio]{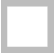} & \includegraphics[width=3mm, keepaspectratio]{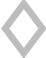} &
    \includegraphics[width=3mm, keepaspectratio]{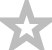}\\
    2.0 & 1.90 & 2.5 & 287 & 0.61 & 492.62 & 18.63 & 8.15 & 0.87 & 4140 & 1.73 & -- & \includegraphics[width=3mm, keepaspectratio]{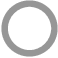} & \includegraphics[width=3mm, keepaspectratio]{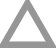} & \includegraphics[width=3mm, keepaspectratio]{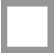} & \includegraphics[width=3mm, keepaspectratio]{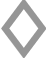} &
    \includegraphics[width=3mm, keepaspectratio]{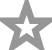}\\ 
    2.0 & 1.90 & 4.5 & 285 & 0.61 & 491.16 & 17.86 & 7.85 & 0.78 & 6933 & 1.60 & -- & \includegraphics[width=3mm, keepaspectratio]{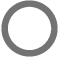} & \includegraphics[width=3mm, keepaspectratio]{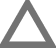} & \includegraphics[width=3mm, keepaspectratio]{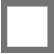} & \includegraphics[width=3mm, keepaspectratio]{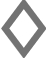} &
    \includegraphics[width=3mm, keepaspectratio]{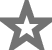}\\
    2.9 & 2.89 & 4.5 & 289 & 0.40 & 602.71 & 26.11 & 7.81 & 0.75 & 2233 & 1.50 & -- & \includegraphics[width=3mm, keepaspectratio]{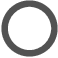} & \includegraphics[width=3mm, keepaspectratio]{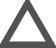} & \includegraphics[width=3mm, keepaspectratio]{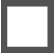} & \includegraphics[width=3mm, keepaspectratio]{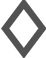} &
    \includegraphics[width=3mm, keepaspectratio]{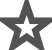} \\
    \hline
    \multicolumn{17}{c}{\textbf{Rough Wall (P60)}} \\
    0.3 & 0.29 & 1.5 & 296 & 0.99 & 97.66 & 4.89 & 16.57 & 1.99  & 7427 & 4.95 & 9.84 & \includegraphics[width=3mm, keepaspectratio]{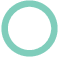} & \includegraphics[width=3mm, keepaspectratio]{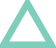} & \includegraphics[width=3mm, keepaspectratio]{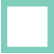} & \includegraphics[width=3mm, keepaspectratio]{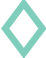}&
    \includegraphics[width=3mm, keepaspectratio]{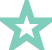}\\
    0.8 & 0.82 & 1.5 & 289 & 0.89 & 261.93 & 13.82 & 24.44 & 2.56 & 22391 & 4.97 & 11.80 & \includegraphics[width=3mm, keepaspectratio]{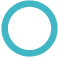} & \includegraphics[width=3mm, keepaspectratio]{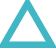} & \includegraphics[width=3mm, keepaspectratio]{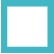} & \includegraphics[width=3mm, keepaspectratio]{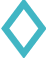}&
    \includegraphics[width=3mm, keepaspectratio]{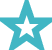}\\
    2.0 & 1.95 & 2.5 & 284 & 0.60 & 496.40 & 34.70 & 11.61 & 1.50 & 10446 & 5.82 & 13.96 & \includegraphics[width=3mm, keepaspectratio]{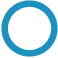} & \includegraphics[width=3mm, keepaspectratio]{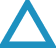} & \includegraphics[width=3mm, keepaspectratio]{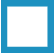} & \includegraphics[width=3mm, keepaspectratio]{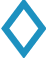}&
    \includegraphics[width=3mm, keepaspectratio]{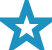}\\
    2.0 & 1.91 & 4.5 & 286 & 0.61 & 492.10 & 34.60 & 11.92 & 1.52 & 20271 & 5.99 & 15.33 & \includegraphics[width=3mm, keepaspectratio]{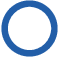} & \includegraphics[width=3mm, keepaspectratio]{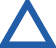} & \includegraphics[width=3mm, keepaspectratio]{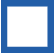} & \includegraphics[width=3mm, keepaspectratio]{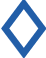}&
    \includegraphics[width=3mm, keepaspectratio]{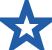}\\
    2.9 & 2.84 & 4.5 & 291 & 0.41 & 601.55 & 62.89 & 10.75 & 1.32 & 7812 & 8.93 & 16.41 & \includegraphics[width=3mm, keepaspectratio]{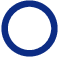} & \includegraphics[width=3mm, keepaspectratio]{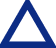} & \includegraphics[width=3mm, keepaspectratio]{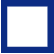} & \includegraphics[width=3mm, keepaspectratio]{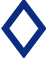}&
    \includegraphics[width=3mm, keepaspectratio]{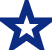}\\
    \hline
    \multicolumn{17}{c}{\textbf{Rough Wall (P24)}} \\
    0.3 & 0.28 & 1.5 & 295 & 0.99 & 97.02 & 5.67 & 18.83 & 2.53 & 9830 & 6.74 & 13.29 & \includegraphics[width=3mm, keepaspectratio]{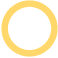} & \includegraphics[width=3mm, keepaspectratio]{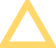} & \includegraphics[width=3mm, keepaspectratio]{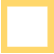} & \includegraphics[width=3mm, keepaspectratio]{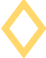} &
    \includegraphics[width=3mm, keepaspectratio]{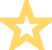}\\
    0.8 & 0.80 & 1.5 & 287 & 0.90 & 255.33 & 16.42 & 24.17 & 2.96 & 27025 & 7.42 & 15.86 & \includegraphics[width=3mm, keepaspectratio]{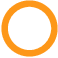} & \includegraphics[width=3mm, keepaspectratio]{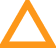} & \includegraphics[width=3mm, keepaspectratio]{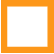} & \includegraphics[width=3mm, keepaspectratio]{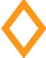}& \includegraphics[width=3mm, keepaspectratio]{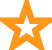}\\
    2.0 & 1.99 & 2.5 & 279 & 0.59 & 497.41 & 41.40 & 14.90 & 2.03 & 15594 & 8.11 & 17.09 & \includegraphics[width=3mm, keepaspectratio]{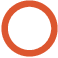} & \includegraphics[width=3mm, keepaspectratio]{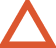} & \includegraphics[width=3mm, keepaspectratio]{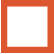} & \includegraphics[width=3mm, keepaspectratio]{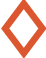} & \includegraphics[width=3mm, keepaspectratio]{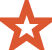}\\
    2.0 & 1.94 & 4.5 & 286 & 0.60 & 496.18 & 42.76 & 15.05 & 2.03 & 30292 & 8.89 & 19.22 & \includegraphics[width=3mm, keepaspectratio]{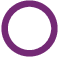} & \includegraphics[width=3mm, keepaspectratio]{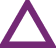} & \includegraphics[width=3mm, keepaspectratio]{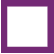}  & \includegraphics[width=3mm, keepaspectratio]{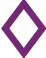} & \includegraphics[width=3mm, keepaspectratio]{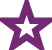}\\
    2.9 & 2.77 & 4.5 & 292 & 0.42 & 596.43 & 67.65 & 12.92 & 1.64 & 11167 & 10.82 & 18.61 & \includegraphics[width=3mm, keepaspectratio]{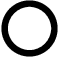} & \includegraphics[width=3mm, keepaspectratio]{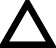} & \includegraphics[width=3mm, keepaspectratio]{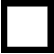} & \includegraphics[width=3mm, keepaspectratio]{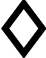}& \includegraphics[width=3mm, keepaspectratio]{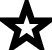}\\
    \hline
    \end{tabular}
    \caption{Summary of all test cases. The air properties are: $M$ (nominal test section Mach number), $M_{\infty}$ (freestream Mach number obtained from PIV measurements at the measurement stations in figure \ref{fig:setup}c), stagnation pressure and temperature ($P_0$ and $T_0$). $T_{\infty}/T_w$ is the freestream-to-wall temperature ratio. The 99\% boundary-layer thickness $\delta$ and momentum thickness $\theta$ are obtained from untransformed velocity profiles extracted either from measurement station \textbf{A} or \textbf{B} in figure \ref{fig:setup}(c). The friction velocity $U_{\tau}$ and other related quantities ($Re_{\tau}$, $C_f$, and $\Delta U_I^+$) are estimated using methods described in \S\ref{sub:wss} all transformations in table \ref{tab:transform}. For brevity, only results from \cite{vandriest1951} transformation are shown here. Symbols denote various transformations (table \ref{tab:transform}), with lighter to darker colour schemes mark the lowest to highest Mach numbers.}
    \label{tab:cases}
    \end{center}
\end{table}

\begin{figure}
    \centering
    \includegraphics[width=13.5cm, keepaspectratio]{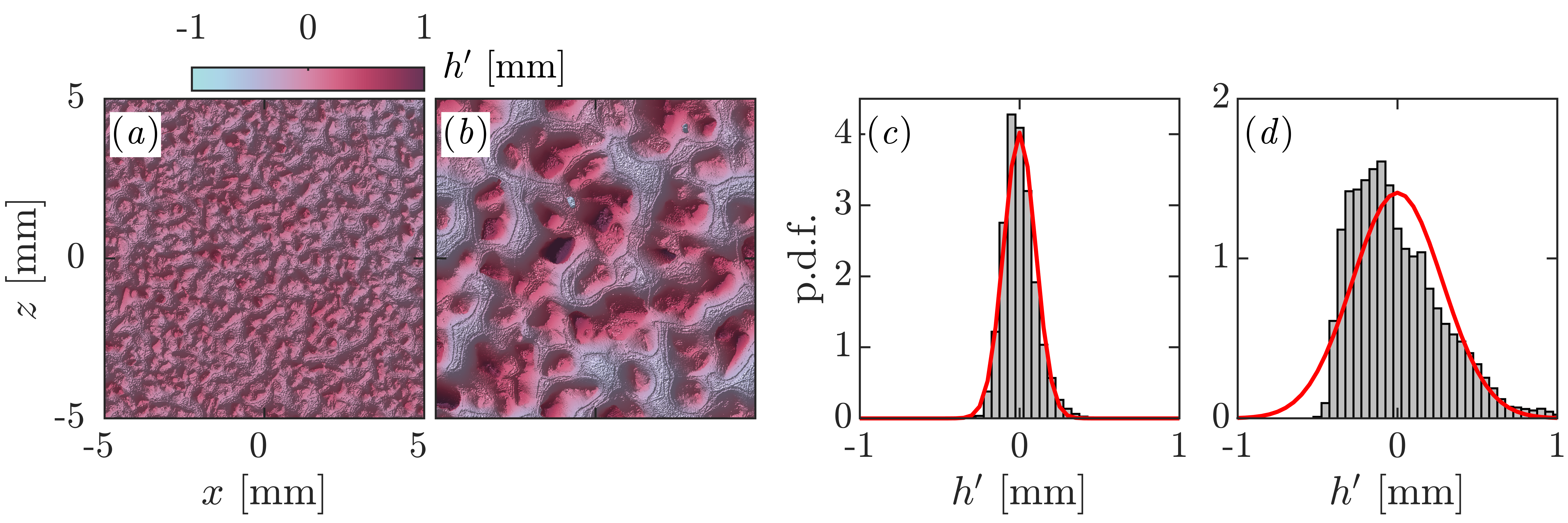}
    \caption{Coloured contours of the scanned sandpaper surface deviation from its mean height, $h' = h - \overline{h}$, shown for 10 mm $\times$ 10 mm coupons of (a) P60- and (b) P24-grit sandpapers. Probability density function (p.d.f.) of $h'$ of the scanned (c) P60- and (d) P24-grit sandpapers. Solid red line (\solidline[red]) is the fitted Gaussian distribution of the p.d.f.}
    \label{fig:sandpaper}
\end{figure}

\begin{table}
    \begin{center}
    \def~{\hphantom{0}}
    \begin{tabular}{l c c c c c c c}
    \hline
    Parameter: & $k$ & $k_a$ & $k_p$ & $k_{rms}$ & $k_{sk}$ & $k_{ku}$ & $ES_x$ \\
    \hline
    Formula: & $6\sqrt{\overline{h'^2}}$ & $\overline{|h'|}$ & $\max(h') - \min(h')$ & $\sqrt{\overline{h'^2}}$ & $\overline{h'^3}/h_{rms}^3$ & $\overline{h'^4}/h_{rms}^4$ & $\overline{\left|\frac{\mathrm{d}h'}{\mathrm{d}x}\right|}$ \\
    Unit:   & mm & mm & mm & mm & -- & -- & -- \\
    \hline
    P60 & 0.595 & 0.077 & 0.804 & 0.099 & 1.261 & 10.375 & 0.541\\
    P24 & 1.695 & 0.226 & 2.023 & 0.283 & 1.841 & 9.708 & 0.659\\
    \hline
    \end{tabular}
    \caption{Key surface parameters from the scanned P60- and P24-grit sandpaper coupons.}
    \label{tab:sandpaper}
    \end{center}
\end{table}

Measurements are conducted for the TBLs developing over a 750 mm $\times$ 300 mm $\times$ 19.5 mm ($58\delta_0 \times 22\delta_0 \times 1.4\delta_0$, length $\times$ width $\times$ maximum height) smooth baseplate (figure \ref{fig:setup}a) suspended at the test section half-height, where $\delta_0$ is the reference 99\% smooth-wall boundary-layer thickness at $M = 0.3$. Flow measurements are conducted on the TBL developing over the flat surface of the baseplate, while the bottom surface has the wedge angles of $\theta_{LE} = 12.6^{\circ}$ and $\theta_{TE} = 15.2^{\circ}$ at the leading and trailing edges, respectively, as illustrated in figure \ref{fig:setup}(b). Eight removable inserts, each has the dimensions of 157.5 mm $\times$ 125 mm, are arranged in a manner shown in figure \ref{fig:setup}(a), allowing multiple changes of test surfaces without removing the base plate from the test section. With this configuration, the rough test surfaces only cover an area of 630 mm $\times$ 250 mm ($42\delta_0 \times 18\delta_0$) of the baseplate, while the edges of the baseplate are smooth (figure \ref{fig:setup}a).    

Test surfaces comprise of a smooth wall (case `SW') and two rough walls constructed from P60-and P24-grit sandpaper sheets (cases `P60' and `P24', respectively). Surface topologies are obtained from samples of P60 and P24 sandpapers prior to the measurement campaigns. A 10 mm $\times$ 10 mm coupon of P60 and a 13 mm $\times$ 13 mm coupon of P24 sandpapers are scanned using an Alicona G4 InfiniteFocus optical profilometer. The topologies are shown as the scanned surface deviations from their mean heights ($h' = h - \overline{h}$) in figures \ref{fig:sandpaper}(a,b), while their roughness height distributions are shown in figures \ref{fig:sandpaper}(c,d). The physical height of the roughness is defined by $k = 6 \sqrt{\overline{h'^2}}$, yielding $k = 0.595$ mm for P60 and $k = 1.695$ mm for P24 sandpapers, respectively. This parameter, alongside other key surface parameters similar to those of \cite{squire2016} and \cite{gul2021}, are presented in table \ref{tab:sandpaper}. The sandpaper sheets are cut to the size of the baseplate inserts (figure \ref{fig:setup}a) and attached to the inserts using double-sided carpet tapes. Since the flow encounters two step changes in roughness as the TBL develops: first the smooth leading edge to the rough surface inserts, followed by the smooth trailing edge of the baseplate, extra care is taken to minimise the height difference between the smooth baseplate and the rough surfaces by lowering the insert's thickness by $\Delta h = $ 1 mm and 1.5 mm for P60 and P24 surfaces, respectively (see inset in figure \ref{fig:setup}a). The required $\Delta h$ values are determined by taking the side images of the P60 and P24 sandpaper coupons (with their base papers and double-sided tapes) against a metal shim of known thickness.

For analyses in this study, velocity profiles are extracted from and averaged over $x = 407 \pm 5$ mm ($x/\delta_0 \approx 29 \pm 0.36$) from the leading edge of the baseplate for SW and P24 cases (station `\textbf{A}' in figure \ref{fig:setup}a) to improve statistical convergence. Due to wall reflections issue in the PIV images, the profiles of P60 cases are instead extracted from and averaged over station `\textbf{B}', $x = 417 \pm 5$ mm ($x/\delta_0 \approx 30 \pm 0.36$, figure \ref{fig:setup}a). Table \ref{tab:cases} shows the complete test matrix and the corresponding test parameters measured at these stations. For brevity, only friction velocity $U_{\tau}$, the Hama roughness function $\Delta U_I^+$, and subsequently the friction Reynolds number $Re_{\tau} = \delta U_{\tau}/\nu_w$ and the skin friction coefficient,
\begin{equation}
    C_f = 2 \frac{\rho_w}{\rho_{\infty}} \left(\frac{U_{\tau}}{U_{\infty}} \right)^2
    \label{eq:Cf}
\end{equation}
estimated from the profiles stretched by VD transformation are shown in table \ref{tab:cases}, denoted by subscript `VD'. Note that, here, $U_{\infty}$ refers to the actual freestream velocity, without any transformation. All test surfaces are subjected to a range of Mach numbers ($0.3 \leq M \leq 2.9$) and friction Reynolds numbers ($4355 \leq Re_{\tau_{VD}} = \delta U_{\tau_{VD}}/\nu_w \leq 30292$). At a constant Mach number ($M = 2$), the stagnation pressure of the tunnel $P_0$ is varied from 2.5 to 4.5 bar (table \ref{tab:cases}), allowing an independent sweep of $Re_{\tau}$ while maintaining a constant $M$. 

\subsection{PIV measurements}
\label{sub:piv}

Non-time-resolved planar 2D2C particle image velocimetry (PIV) measurements are conducted on the streamwise--wall-normal ($x$--$y$) plane, shown in figure \ref{fig:setup}(b,c), located at $z = 12$ mm from the centreline of the tunnel (figure \ref{fig:setup}a). The measurement plane is illuminated by a laser sheet generated by an InnoLas SpitLight Compact PIV 400 dual-pulse Nd:YAG laser with a typical optical configuration. The flow is seeded with DEHS (Di-Ethyl-Hexyl-Sebacat) tracer particles, whose mean diameter is approximately 1 $\upmu$m \citep{kahler2002}. As many as 400 image pairs are captured for each test case shown in table \ref{tab:cases} by two 16-bit, 5.5 MP LaVision Imager sCMOS CLHS cameras ($2560 \times 2160$ pixels sensor resolution), each equipped with a Zeiss Milvus 100mm f/2.0 lens. White dashed lines in figure \ref{fig:setup}(c) mark the FOV (field-of-view) captured by each camera, with 10-mm overlap between the cameras. Stitched together, the camera arrangement yields a FOV spanning 109 mm in length and 32 mm in height ($7.8 \delta_0 \times 2.3 \delta_0$). 

For all test cases in table \ref{tab:cases}, PIV images are acquired at a constant rate of $f = 10.5$ Hz. The time between images $\Delta t$ is set to be inversely proportional to the nominal test section Mach numbers, between 0.65 to 4 $\upmu$s for test cases $M = 2.9$ to $M = 0.3$, respectively. For the reference test case (SW at $M = 0.3$), this acquisition setup corresponds to the viscous-scaled sampling interval of $t^+ \equiv \Delta t U_{\tau}^2/\nu_w = 2.8$ and the boundary-layer turnover rate of $U_{\infty}/(f\delta_0) = 723$. 

PIV calibration is conducted prior to the measurements. The current setup yields an image pixel size of 27-28 $\upmu$m/pixel between the two cameras. PIV cross-correlation is performed using an in-house PIV package developed at the University of Southampton \citep{lawson2016}. The final correlation window size is $16 \times 16$ pixels (50\% overlap), corresponding to a viscous-scaled spatial resolution of $\Delta x^+ = \Delta y^+ = 134$ for the reference test case. The uncertainties of the PIV measurements are estimated using the nominal subpixel uncertainty of 0.1 pixels \citep{adrian2011}, yielding the uncertainties of $\pm 0.6\%$ and $\pm 3.8\%$ of the maximum pixel displacements for $u$ and $v$ velocity components, respectively ($\pm 1.2\%$ and $\pm 7.6\%$ for their variances).

\subsection{Temperature profiles}
\label{sub:temperature}

\begin{figure}
    \centering
    \includegraphics[width=13.5cm, keepaspectratio]{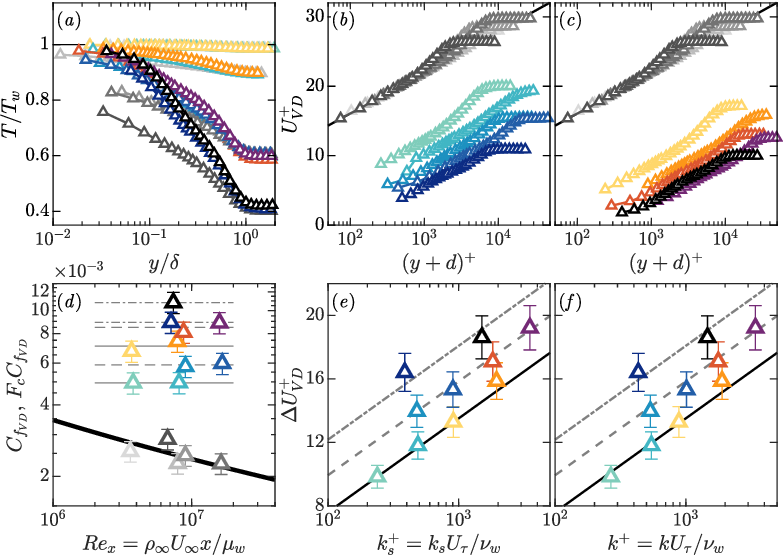}
    \caption{(a) Mean temperature profile $T/T_w$ as function of $y/\delta$ (colours correspond to various test surfaces and $M$ sweep shown in table \ref{tab:cases}). Inner-scaled mean streamwise velocity profile for cases (b) SW and P60, (c) SW and P24 stretched by VD transformation (table \ref{tab:transform}). \solidline: $1/\kappa \log y^+ + B$. (d) Skin friction coefficients $C_f$ vs $Re_x$ estimated from VD-transformed profiles. SW data are multiplied by $F_c$ (equation \ref{eq:hopkins}), \solidline: $[2\log_{10} Re_x - 0.65]^{-7/3}$ \citep{schlichting1960}. Lines correspond to constant $C_f$, \solidline[gray]: $M < 1$, \dashedline[gray]: $M = 2$, and \dashdotline[gray]: $M = 2.9$. $\Delta U_{VD}^+$ as a function of (e) $k_s^+$ and (f) $k^+$ for all rough-wall cases. Error bars in (d--f) denote the uncertainties estimated in \S\ref{sub:mean}. Lines correspond to log-relation, \solidline[black]: $M < 1$, $1/\kappa \log{k_s^+} + B - B_{FR}$, \dashedline[gray]: $M = 2$, and \dashdotline[gray]: $M = 2.9$. Colours corresponds to the test surfaces, $M$ and $Re$ sweeps (table \ref{tab:cases}).}
    \label{fig:mean}
\end{figure}

Due to limitations of the test facility, the mean wall temperature $T_w$ and the mean temperature profile across the boundary layer $T(y)$ are not measured during the experimental campaign. Thus, these quantities are estimated based on similar experiments on rough-wall TBLs conducted in similar test facilities (intermittent, blow-down supersonic wind tunnels). In \cite{goddard1959,latin2000,ekoto2008}, for example, it was assumed that zero-pressure gradient, rough-wall TBLs develop over adiabatic walls. 

Using the same assumption, $T_w$ is equal to the adiabatic wall temperature $T_{aw}$ and the temperature-velocity profile has a quadratic relation to the mean streamwise velocity profile \citep{crocco1948,walz1959} given by,
\begin{align}
     T_w &= T_{aw} = T_{\infty} \left( 1+ r \frac{\gamma-1}{2} M_{\infty}^2 \right) \\
    \frac{T}{T_{\infty}} &= 1 + r \frac{\gamma-1}{2} M_{\infty}^2 \left[ 1 - \left( \frac{U}{U_{\infty}} \right)^2 \right] \label{eq:crocco}
\end{align}
where the recovery factor $r = Pr^{1/3}$ and $Pr = 0.72$ is the Prandtl number \citep{pirozzoli2011}. The mean temperature profiles relative to the wall temperature  $T/T_w$ of all test cases are shown in figure \ref{fig:mean}(a) as a function of wall-normal locations $y/\delta$ (note that $y$ shown in this figure is corrected using techniques described in \S\ref{sub:wss}). At the wall, $T/T_w = 1$ (solid black line in figure \ref{fig:mean}a), and it decreases towards the freestream following the quadratic relation in equation \eqref{eq:crocco}. The freestream-to-wall temperature ratios $T_{\infty}/T_w$ for all test cases are listed in table \ref{tab:cases}, showing that the temperature profile proportionally decreases with $M$. At the lowest Mach number ($M = 0.3$), the compressibility effect is largely negligible as $T_{\infty}$ is 1\% lower than $T_w$  ($T_{\infty}/T_w = 0.99$). As $M$ increases to the transonic flow regime, $T_{\infty}/T_w \approx 0.90$ and finally, in the supersonic regime, it decreases further to $T_{\infty}/T_w \approx 0.60$ and 0.40 at $M = 2$ and 2.9, respectively.     

To account for the compressibility effect, the transformation functions $f_I$ and $g_I$ listed in table \ref{tab:transform} require the density and viscosity profiles relative to the wall, $\rho^+$ and $\mu^+$, to be known. These quantities are derived from the temperature profiles $T^+$, from the boundary layer equation $\partial P/\partial y = 0$ and Sutherland's formula,
\begin{equation}
    \rho^+ = \frac{\rho}{\rho_w} = \left(\frac{T}{T_w}\right)^{-1} \quad \mathrm{and} \quad \mu^+ = \frac{\mu}{\mu_w} = \left(\frac{T}{T_w}\right)^{1.5} \left( \frac{T_w + S}{T + S} \right) 
    \label{eq:sutherland}
\end{equation}
where $S = 110.4$ K.

\subsection{Wall position correction and shear stress estimation}
\label{sub:wss}

Similar to the temperature profiles and wall temperatures, the wall shear stress (WSS) is not directly measured in the experiment. For all test cases in table \ref{tab:cases}, a modified \cite{clauser1954} fitting is employed. This method requires some correction of the wall-normal coordinates alongside the friction velocity $U_{\tau}$ estimation. Thus, to minimise uncertainty, another technique is applied to determine the wall location \emph{before} estimating $U_{\tau}$ with Clauser fitting. 

For SW test cases, the wall-normal grid position $y$ is first corrected based on the location of the light-sheet reflection on the wall in the PIV images. Then, the Clauser method searches for a set of parameters $(U_{\tau},\varepsilon)$, where $\varepsilon$ is a further wall correction of a much smaller order ($O$(0.1 mm)) than the first wall correction determined using PIV images, that minimises the r.m.s. (root mean square) error of  
\begin{equation}
\Psi_{smooth} = U_I^+ - \left[ \frac{1}{\kappa} \log{ \left( \frac{(y+\varepsilon)_I U_{\tau}}{\nu_w}\right)} + B \right]    
\end{equation}
in the log region $30 \leq (y+\varepsilon)_I^+ \leq 0.15 Re_{\tau_I}$, where $Re_{\tau_I} = \delta_I U_{\tau}/\nu_w$ and $\delta_I$ is the boundary-layer thickness of the transformed velocity profile. 

For P60 and P24 cases, two parameters are added to the set: the zero plane displacement $d$ and the transformed Hama roughness function $\Delta U_I^+$ (equation \ref{eq:rough}). Similar to the approach by \cite{squire2016,gul2021} for sandpaper roughness, as well as \cite{modesti2022} for cube roughness (figure \ref{fig:roughness_type}\textbf{D}), $d$ is kept constant at $d = -k/2$, where the physical roughness height $k$ is determined through surface scanning (table \ref{tab:sandpaper}). The sandpaper grain reflection on the PIV images is then used to determine the wall location and its downward shift by $k/2$ for the rough wall cases. The final parameter set is $(U_{\tau},\varepsilon,\Delta U_I^+)$ that minimise the r.m.s. error of 
\begin{equation}
\Psi_{rough} = U_I^+ - \left[ \frac{1}{\kappa} \log{ \left( \frac{(y - k/2 + \varepsilon)_I U_{\tau}}{\nu_w}\right)} + B - \Delta U_I^+ \right] 
\end{equation}
within the inertial range $3\sqrt{Re_{\tau_I}} \leq (y-k/2+\varepsilon)_I^+ \leq 0.15 Re_{\tau_I}$ \citep{squire2016,gul2021}. To further minimise the uncertainty for rough wall cases, $\varepsilon$ is only fitted for profiles stretched by the VD transformation, where $y = y_I$ (table \ref{tab:transform}). This wall correction $\varepsilon_{VD}$ is then applied to the profiles before stretching them with other transformations listed in table \ref{tab:transform}, and the wall location $y$ is not corrected further in the Clauser fitting.    

\section{Mean flow and rough-wall skin friction}
\label{sub:mean}

The inner-scaled transformed mean streamwise velocity $U_I^+ = U_I/U_{\tau}$ profiles for the P60 and P24 rough-wall cases are shown in figure \ref{fig:mean}(b,c), respectively, as a function of the inner-scaled transformed wall-normal coordinates $(y+d)_I^+ = (y+d)_I U_{\tau}/\nu_w$. For brevity, here we will only discuss the VD-transformed profiles, $U_{VD}^+$, and sensitivity to using a different transformation will be discussed in the following section. Gray colour schemes denote SW cases (table \ref{tab:cases}), showing the transformed profiles collapsing onto the log relation $1/\kappa \log y^+ + B$, as prescribed by the modified Clauser fitting in \S\ref{sub:wss}. In contrast, the rough wall cases show a clear downward vertical shift by $\Delta U_{VD}^+$, obtained from \S\ref{sub:wss}.

From the fitted $U_{\tau}$, the skin friction coefficient $C_f$ is calculated using equation \eqref{eq:Cf}. Figure \ref{fig:mean}(d) shows $C_f$ as a function of Reynolds number based on the fetch lengths $Re_x$ for all test cases. The uncertainties of $C_f$ are estimated via the propagation of uncertainty analysis of equation \eqref{eq:Cf}. Here, we assume the expected $U_{\tau}$ error range of $\pm 5\%$ for the current estimation method in \S\ref{sub:wss} \citep{squire2016}. Other variable, $\rho_w/\rho_{\infty}$, is a function of $M_{\infty}$ via equations \eqref{eq:sutherland} and \eqref{eq:crocco}. The uncertainty of $U_{\infty}$ is estimated via the uncertainty of the maximum pixel displacement in $x$ of $\pm 0.6\%$ (outlined in \S\ref{sub:piv}). The propagation of uncertainty analysis yields the $C_f$ estimation error of $\pm 10.28\%$, denoted by the error bars in figure \ref{fig:mean}(d). The black solid line denotes Schlichting's fit for incompressible, smooth-wall TBLs, $C_f = [2 \log_{10} Re_x - 0.65]^{-7/3}$ \citep{schlichting1960}. To match the incompressible smooth-wall $C_f$, \cite{hopkins1971} derived a correction factor $F_c$ from the VD II transformation \citep{vandriest1956} for adiabatic walls,
\begin{equation}
    F_c = \frac{T_w/T_{\infty} - 1}{\arcsin^2 \alpha}, \quad \mathrm{where} \quad \alpha = \frac{T_w/T_{\infty}-1}{\sqrt{T_w/T_{\infty}(T_w/T_{\infty}-1)}}
    \label{eq:hopkins}
\end{equation}
such that the corrected skin friction of a compressible smooth wall $F_c C_f$ is equivalent to that of the incompressible flow. It is noted that $F_c$ defined in equation \eqref{eq:hopkins} is also equal to $(\rho_{\infty}/\rho_w)(U_{\infty}/U_{\infty_{VD}})^2$, where $U_{\infty_{VD}}$ is the freestream velocity of the VD-transformed velocity profile. Meanwhile, the Reynolds number is corrected by a factor of $\mu_{\infty}/\mu_w$ \citep{hopkins1971}, which results in the definition of $Re_x = \rho_{\infty} U_{\infty} x/\mu_w$ as the $x$-axis of figure \ref{fig:mean}(d). Grey coloured symbols in figure \ref{fig:mean}(d) are $F_c C_f$ as a function of $Re_x$ for SW cases, showing that the estimated (in \S\ref{sub:wss}) and corrected $C_f$ across $M$ and $Re$ sweeps largely match Schlichting's fit for incompressible, smooth-wall TBLs.  

To the best of the authors' knowledge, no correction factor exists for $C_f$ of compressible, rough-wall TBLs. Therefore, the estimated $C_f$ for P60 and P24 cases in figure \ref{fig:mean}(d) are presented without any correction factor, following \cite{goddard1959}. For all rough-wall cases, the fitted $\Delta U_I^+ > 7$, and thus these cases are within the fully rough regime. This is evident in the subsonic to transonic cases ($M = 0.3$ and 0.8): $C_f$ is independent of $Re_x$ for the same sandpaper roughness and the data largely collapse to the grey solid lines in figure \ref{fig:mean}(d). As the flow reaches the supersonic regime, it appears that there is an additional dependency on $M$. For the same roughness (either P60 or P24), $M = 2$ and 2.9 cases suggest a higher $C_f$ from that of subsonic and transonic cases, which increases proportionally with $M$ (gray dashed and dash-dot lines in figure \ref{fig:mean}d). At $M = 2$, where $Re$ sweep is available (table \ref{tab:cases}), $C_f$ appears to be independent of $Re_x$, as suggested by the estimated $C_f$ data collapsing onto grey dashed lines in figure \ref{fig:mean}(d). Figures \ref{fig:mean}(e,f) show the fitted Hama roughness functions from the stretched velocity profiles $\Delta U_I^+$ (in \S\ref{sub:wss}) for all P60 and P24 cases (table \ref{tab:cases}). The error bars in figures \ref{fig:mean}(e,f) show the uncertainties of $\Delta U_{VD}^+ \pm 7.3\%$ for all test cases listed in table \ref{tab:cases}, based on propagating the $C_f$ uncertainties via equation \eqref{eq:hama_cf}.


Since we are looking to transform the established framework for incompressible, rough-wall TBLs to the compressible one (figure \ref{fig:framework}), $k_s$ is determined solely from the fitted $\Delta U_{VD}^+$ for the incompressible cases at $M = 0.3$. The P60 and P24 cases at $M = 0.3$ are within the fully rough regime, and thus the incompressible $k_s$ can be derived directly from equation \eqref{eq:hama}, yielding $k_s =$ 0.531 and 1.742 mm for P60 and P24 cases, respectively. In figure \ref{fig:mean}(e), `$k_s$' refers to these values.

Using the above incompressible $k_s$, the fitted $\Delta U_{VD}^+$ for all P60 and P24 cases in table \ref{tab:cases} is shown in figure \ref{fig:mean}(e) as a function of $k_s^+ = k_s U_{\tau}/\nu_w$. Results suggest that all cases are within the fully rough regime ($\Delta U_{VD}^+ > 7$ or $k_s^+ > 70$). The collapse of $M = 0.3$ data to the log relation (equation \eqref{eq:hama}, black solid line in figure \ref{fig:mean}e) is given since the incompressible $k_s$ is derived from this relation. However, the collapse is also observed for the transonic ($M = 0.8$) cases. As $M$ reaches the supersonic regime, there seems to be a proportional relation between $M$ and the log-law intercept: the intercept $B-B_{FR}$ (equation \ref{eq:hama}) increases with $M$ for the supersonic cases (grey dashed line for $M = 2$ and grey dash-dot line for $M = 2.9$ in figure \ref{fig:mean}e), possibly related to the wave drag generated by protrusion of the roughness elements. These observations still hold if the fitted $\Delta U_{VD}^+$ is shown as a function of the inner-scaled physical roughness height $k^+ = k U_{\tau}/\nu_w$ instead of $k_s^+$, as shown in figure \ref{fig:mean}(f). 

\section{Correction factors}
\label{sub:corr_factors}

\begin{figure}
    \centering
    \includegraphics[width=13.2cm, keepaspectratio]{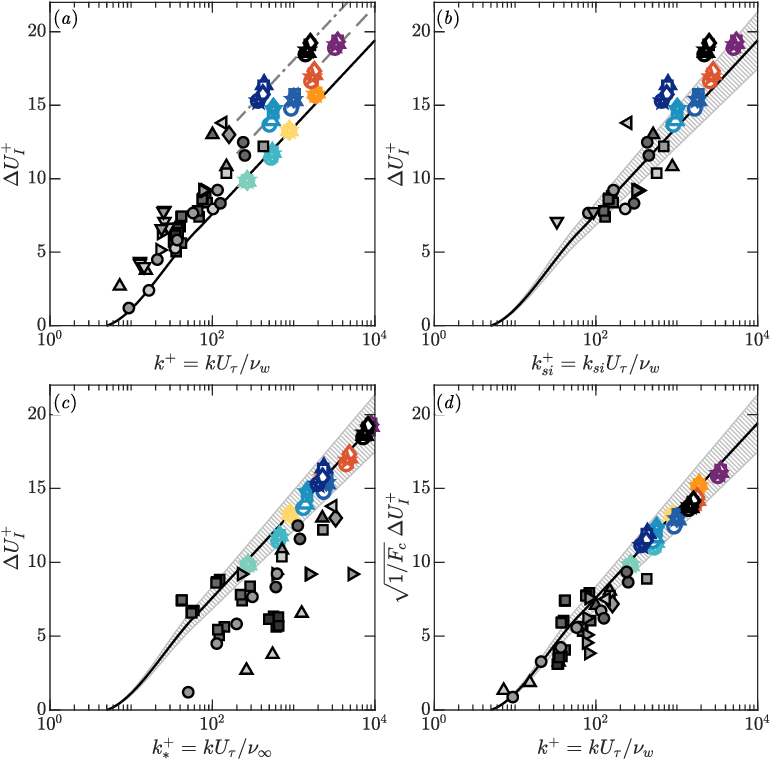}
    \caption{Hama roughness function (obtained from transformed velocity profiles) $\Delta U_I^+$ of all test cases in table \ref{tab:cases} and previous studies in table \ref{tab:ref} as a function of: (a) $k^+ = k U_{\tau}/\nu_w$, (b) $k_{si}^+ = k_{si} U_{\tau}/\nu_w$ ($k_{si}$ is obtained from matched studies in incompressible flow regime, as outlined in \S\ref{sub:match}), and (c) $k_{\ast}^+ = k U_{\tau}/\nu_{\infty}$. In (d), $\Delta U_I^+$ is multiplied by $\sqrt{1/F_c}$ (see equation \ref{eq:hopkins} for the definition of $F_c$). \solidline: Nikuradse's data \citep{nikuradse1933}, with $\kappa = 0.39$, $B = 4.3$, and $B_{FR} = 8.5$. In (a),  the intercept $B-B_{FR}$ are shown by \dashedline[gray]: $M = 2$, and \dashdotline[gray]: $M = 2.9$. In (b--d), hatched regions mark the $\pm$10\% error from Nikuradse's data. Open coloured symbols correspond to test cases in table \ref{tab:cases}, while gray-filled symbols correspond to the previous studies in table \ref{tab:ref}.}
    \label{fig:hama}
\end{figure}

Figure \ref{fig:hama}(a) summarises the estimated Hama roughness function $\Delta U_I^+$ for the test cases in table \ref{tab:cases} and previous studies (in gray filled symbols) listed in table \ref{tab:ref} for various roughness families in figure \ref{fig:roughness_type}. Various open symbols in figure \ref{fig:hama}(a) correspond to the fitted $\Delta U_I^+$ of the present test cases in table \ref{tab:cases}, obtained from velocity profiles stretched by various transformations listed in table \ref{tab:transform}. For the current experimental data, the fitted $\Delta U_I^+$ is relatively independent of transformations to account for compressibility effects. The same phenomenon observed for $\Delta U_{VD}^+$ in figure \ref{fig:mean}(f) is still apparent if different transformations are used to obtain $\Delta U_I^+$: the collapse of subsonic data ($M \leq 0.8$) to Nikuradse's formula for incompressible flows (equation \ref{eq:hama}) and the subsequent upward shift of the log-law intercept that seems to be proportional to $M$ at supersonic regime (gray dashed and dash-dot lines in figure \ref{fig:hama}a). 

As in the current study, previous studies estimated $\Delta U_I^+$ by fitting the downward shift of the transformed velocity profiles. The friction drag was either obtained from numerical simulations, directly measured, or, when direct measurement was not possible in the facilities, a fitting method (similar to that described in \S\ref{sub:wss}), as outlined in \S\ref{sub:intro}. In terms of $k_s$, it is worth noting that: (i) the collapse to the log-relation in equation \eqref{eq:hama} is forced since $k_s$ was derived directly from the estimated $\Delta U_I^+$ for a given $M$ and $Re$ in the majority of the previous studies, (ii) $k_s$ was not reported in some studies \citep{goddard1959,reda1975,sahoo2009,tyson2013,aghaei-jouybari2023,wang2024}, and (iii) the present experimental data (figures \ref{fig:mean}e,f) show that there is only a slight difference between using the incompressible $k_s$ or the physical roughness height $k$ to scale $\Delta U_I^+$. Considering these factors, figure \ref{fig:hama}(a) scales the $x$-axis with $k$ instead of $k_s$, and the subsequent analyses in this study utilise $k$ instead of the reported $k_s$.

Observing previous studies' data in figure \ref{fig:hama}(a), it is apparent that the majority of the test cases do not collapse to Nikuradse's data \citep{nikuradse1933} for either hydrodynamically smooth ($k_s^+ < 5$), transitionally ($5 \leq k_s^+ \leq 70$), or fully rough surfaces ($k_s^+ > 70$). To be consistent with present data, we use $\kappa = 0.39$, $B = 4.3$, and $B_{FR} = 8.5$. There appears to be a tendency towards an upward shift in the intercept $B-B_{FR}$, similar to that observed in the current study. Although the shift is likely caused by the $x$-axis scaling by $k^+$ instead of $k_s^+$, a collapse to Nikuradse's data is observed for some sandpaper roughness cases \citep{goddard1959,reda1975,latin2000} and some cube roughness cases \citep{modesti2022}. Whether this upward shift is proportional to $M$, as suggested by the current experimental data, is even more difficult to say as the test cases in the previous studies covered 7 different roughness families (figure \ref{fig:roughness_type}), each with variations of their roughness parameters, with the majority of studies conducted at a single $M$ and $Re$. 

Taking the current and previous studies' data together, it is also apparent that in order to utilise the incompressible flow framework for drag estimation in the compressible flow regime (figure \ref{fig:framework}), either $k$ (or $k_s$), i.e. the $x$-axis in figure \ref{fig:hama}(a), or the $\Delta U_I^+$ (the $y$-axis in the same figure) must be scaled by a factor to account for the wave drag effect. Two possible scaling factors will be explored and discussed in this section.    

\subsection{Defining equivalent incompressible test cases}
\label{sub:match}

\afterpage{
\clearpage
\begin{landscape}
    \centering
    \includegraphics[width=12cm, keepaspectratio]{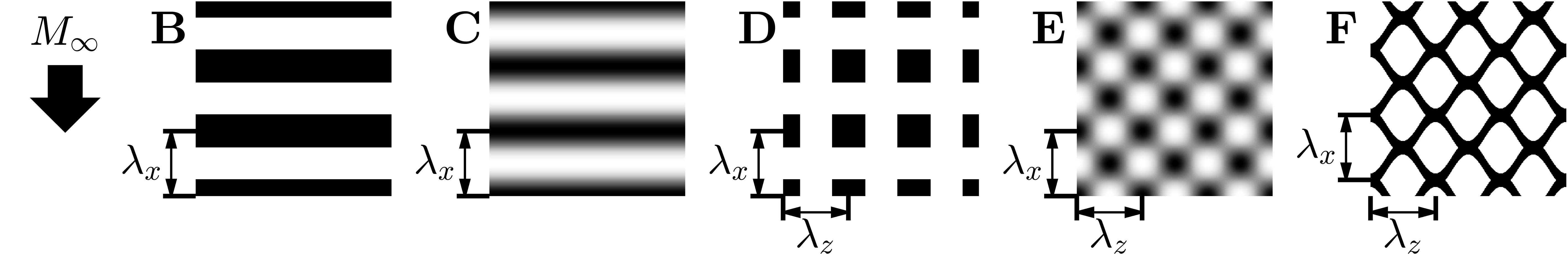}
    
    \def~{\hphantom{0}}
    \begin{tabular}{c l c c c c c l c c c c c c}
    \hline
    Type & Compressible flow ($M > 1$) & $\delta/k$ & $\lambda_x/\delta$ & $\lambda_z/\delta$ & $Re_{\tau}$ & $\Delta U_I^+$ & Incompressible flow ($M \approx 0$) & $\delta/k$ & $\lambda_x/\delta$ & $\lambda_z/\delta$ & $Re_{\tau}$ & $\Delta U^+$ & $k_s/k$ \\
    \hline
     \multirow{7}{*}{\textbf{A}} & Present study, P60 & $\approx$19 & -- & -- & $\approx$12843 & $\approx$15 & \cite{flack2007} & 21 & -- & -- & 4140 & 10.8 & 1.80 \\
     & Present study, P24 & $\approx$8.4 & -- & -- & $\approx$19017 & $\approx$18 & \cite{castro2007}$^\dagger$ & 9.6 & 0.21 & 0.21 & 3867 & 12.2 & 1.56 \\
     & \cite{goddard1959} & 15.3 & -- & -- & -- & 7.9 & \cite{flack2007} & 16 & -- & -- & 6060 & 13 & 2.26 \\
     & \cite{reda1975} & 40.2 & -- & -- & -- & 7.7 & \cite{gul2021} & 41.9 & -- & -- & 2275 & 6.9 & 1.40 \\
     & \cite{reda1975} & 43.1 & -- & -- & -- & 9.2 & \cite{gul2021} & 44.2 & -- & -- & 3276 & 7.8 & 1.42 \\
     & \cite{latin2000} & 27.7 & -- & -- & 3474 & 8.3 & \cite{brzek2007} & 25 & -- & -- & 3789 & 9.7 & 2.36 \\
     & \cite{latin2000} & $\approx$20 & -- & -- & $\approx$5056 & $\approx$12 & \cite{flack2007} & 21 & -- & -- & 4140 & 10.8 & 1.80 \\
     \hline
     
     \multirow{2}{*}{\textbf{B}} & \cite{latin2000} & 30 & 0.13 & -- & 4454 & 10.8 & \cite{krogstad2005} & 29.4 & 0.27 & -- & 600 & 8.7 & 5.93\\
     & \cite{sahoo2009} & 7.3 & 0.68 & -- & 733 & 13 & \cite{leonardi2003} & 5 & 1 & -- & 460 & 11.7 & 5.09 \\
     \hline
     
     \multirow{2}{*}{\textbf{C}} & \cite{tyson2013} & 6.3 & 2 & -- & 160 & 7.1 & \cite{tyson2013} & 6.3 & 2 & -- & 160 & 7.8 & 1.28\\
     & \cite{aghaei-jouybari2023} & 10 & 2 & -- & 250 & 7.8 & \cite{hudson1996} & 10 & 2 & -- & 306 & 8.1 & 3.72\\
     \hline
     
     \multirow{3}{*}{\textbf{D}} & \cite{latin2000} & 28.2 & 0.14 & 0.14 & 4184 & 10.4 & \cite{volino2011} & 27.8 & 0.29 & 0.07 & 1869 & 10 & 1.62 \\
     & \cite{ekoto2008} & 14.7 & 0.27 & 0.27 & 6235 & 12.2 & \cite{castro2007} & 17.6 & 0.11 & 0.11 & 3278 & 10.4 & 3.80 \\
     & \cite{modesti2022} & 12.5 & 0.24 & 0.24 & $\approx$1895 & $\approx$8.3 & \cite{modesti2022} & 12.5 & 0.24 & 0.24 & 513 & 7.8 & 1.90 \\
     \hline
     
     \textbf{E} & \cite{wang2024} & $\approx$13 & 0.45 & 0.45 & 608 & 9.2 & \cite{chan2015} & 14 & 0.51 & 0.51 & 540 & 9.5 & 4.10 \\
     \hline
     
     \textbf{F} & \cite{sahoo2009} & 6.3 & 0.28 & 0.73 & 845 & 13.8 & \cite{castro2007} & 9.8 & 0.33 & 0.97 & 978 & 9.2 & 1.84 \\
     \hline
    \end{tabular}
    \captionof{table}{Summary of matched incompressible and compressible rough-wall TBL cases. $\dagger$: matched roughness type not found.}
    \label{tab:match}
\end{landscape}
\clearpage}

In this section, we are taking the first steps of transferring the existing framework for drag characterisation in incompressible flows for compressible flows (figure \ref{fig:framework}). Since most of the studies listed in table \ref{tab:ref} were conducted only in the compressible flow regime, ``incompressible equivalent" test cases for these studies must first be defined. Following this, an attempt is made to match the current test cases (table \ref{tab:cases}) and all the cases from previous studies (table \ref{tab:ref}) with their incompressible equivalents. Assuming that the drag (and the additional wave drag in the case of compressible flows) are proportional to the size of the physical roughness, each compressible ($M_{\infty} > 1$ or $M_b > 1$) case in table \ref{tab:cases} and table \ref{tab:ref} is matched with a previous study of the same roughness family (figure \ref{fig:roughness_type}) and the closest available $\delta/k$ in the incompressible flow regime ($M \approx 0$). Here, we have chosen to match $\delta/k$ rather than $k$, since the roughness elements considered here occupy a significant portion of the log layer (i.e. $\delta/k < 40$), and therefore the equivalent sand gain roughness that fulfills equation \eqref{eq:hama} is expected show a dependency on $k/\delta$ \citep{jimenez2004}. One consequence of this choice is that it is more likely that a sandpaper-type roughness (figure \ref{fig:roughness_type}\textbf{A}) is matched with another study of a different sandpaper grit. Additionally, only cases in the fully rough regime ($\Delta U_I^+ \geq 7$) are considered. Whenever possible, the roughness geometry (i.e. the streamwise and spanwise spacing or the extent of the roughness elements, $\lambda_x$ and $\lambda_z$ illustrated in table \ref{tab:match}) and some flow-related properties, $Re_{\tau}$ and $\Delta U_I^+$ (for $M > 1$, $\Delta U^+$ if $M \approx 0$), are also matched.  

Table \ref{tab:match} lists the matched compressible and incompressible test cases. The incompressible equivalents are found for all roughness families in figure \ref{fig:roughness_type}, except for the diamond-type roughness (figure \ref{fig:roughness_type}\textbf{G}). For every matched roughness family, the closest matching $\delta/k$ case exists, except for the present P24 test cases. As there is no incompressible sandpaper test case that can match $\delta/k \approx 8.4$, a cube roughness (figure \ref{fig:roughness_type}\textbf{D}) of \cite{castro2007} ($\delta/k \approx 9.6$) is chosen as the incompressible equivalent of the P24 test cases. While the majority of the previous studies in table \ref{tab:ref} must be matched with other studies in the incompressible flow regime, for the sine wave roughness (figure \ref{fig:roughness_type}\textbf{C}) of \cite{tyson2013} and the cube roughness of \cite{modesti2022}, the incompressible equivalents are taken from the simulations of the same rough surfaces at $M_b = 0.3$. Since $k_s$ is not reported by \cite{tyson2013}, it is derived from the reported $\Delta U^+$ via equation \eqref{eq:hama}. The reported $k_s/k$ (table \ref{tab:match}) of the incompressible test cases are then utilised to find the equivalent sand grain roughness $k_{si}$ for the matched compressible test cases,
\begin{equation}
    k_{si} = \left. \frac{k_s}{k} \right|_{M \approx 0} \left. k\right|_{M > 1}
\end{equation}

Figure \ref{fig:hama}(b) shows the reported $\Delta U_I^+$ for the present and previous test cases as a function of $k_{si}^+ = k_{si} U_{\tau}/\nu_w$. Black solid line represents Nikuradse's formula for rough walls in incompressible flow regime, while the hatched region is the $\pm 10\%$ deviation from the formulation. Coloured open and gray filled symbols are the data from current and previous compressible flow studies, respectively, with the symbols corresponding to various studies in table \ref{tab:ref} and \ref{tab:cases}. For the current dataset, $k_{si}$ obtained from the matched incompressible test cases (table \ref{tab:match}) collapses $\Delta U_I^+$ to Nikuradse's data for P60 and P24 test cases at $M = 2$, for both lower and higher $Re$ (table \ref{tab:cases}), although notably less so for the P24 test surface at the higher $Re$ ($Re_{\tau} = 30292$, open purple symbols in figure \ref{fig:hama}b). On the other hand, $k_{si}$ fails to collapse $\Delta U_I^+$ for $M = 2.9$ test cases, suggesting that $k_s$, to an extent, is a function of $M$, which has also been suggested if $k_s$ from the $M = 0.3$ data is instead utilised (figure \ref{fig:mean}e,f). However, the estimated $k_{si}$ for the cube roughness of \cite{modesti2022} collapses the reported $\Delta U_I^+$ of the fully rough test cases, across the Mach number and $Re$ sweeps ($M = 2$ and 4, $1034 \leq Re_{\tau} \leq 3659$), within $\pm 10\%$ error from Nikuradse's formula. This suggests the possibility of other criteria to consider for finding $k_{si}$ beyond matching roughness type and $\delta/k$.

First, we consider the roughness geometry as a possible additional criterion for $k_{si}$ determination. For the 2D traverse bars (figure \ref{fig:roughness_type}\textbf{B}) of \cite{sahoo2009}, whose $\lambda/k = 5$ is identical to those of \cite{leonardi2003}, the $k_{si}$ obtained from the latter study collapses the $\Delta U_I^+$. However, $k_{si}$ fails to collapse the $\Delta U_I^+$ of a 2D sine waves of \cite{tyson2013}, although the roughness geometry was identical to that simulated at $M = 0.3$. And yet, $k_{si}$ from the cube roughness of \cite{castro2007} predicts the drag of the same roughness type of \cite{ekoto2008} relatively well, even though the roughness element spacing in the first study is less than half of the latter ($\lambda_x/\delta = \lambda_z/\delta = 0.11$ and 0.27, respectively).     

The second additional criterion to consider is matched $Re$. Data from previous studies seem to suggest that this is not a determining factor: the obtained $k_{si}$ for the P24-grit sandpaper of \cite{reda1975}, as well as P36- and P20-grit sandpapers of \cite{latin2000}, are able to collapse the $\Delta U_I^+$ even though the $Re_{\tau}$ do not precisely match their incompressible equivalents. \cite{latin2000}, for example, conducted the studies at $Re_{\tau} \approx 5000$, while \cite{flack2007} at $Re_{\tau} \approx 4000$. A closer match of $Re_{\tau} \approx 3000$ between the sandpapers of \cite{latin2000} and \cite{brzek2007}, as well as the matched $\Delta U_I^+$ of the cube roughness of \cite{latin2000} and $\Delta U^+$ of \cite{volino2011}, on the other hand, fail to obtain an accurate $k_{si}$ for the aforementioned test surfaces at compressible flow regime. Taken together, these observations from present and previous studies seem to suggest that there is no distinct criterion that can guarantee the success of using $k_{si}$ from available incompressible flow data to predict the drag of rough walls in their equivalent compressible flow test cases -- neither the roughness-related criteria (family type, physical height, and geometry) nor the flow-related criteria ($Re_{\tau}$ or $\Delta U^+$). This leads to another possible correction factor of $k$ (or $k_s$ from incompressible flow data), by considering some compressibility effects within the flows.  

\subsection{Scaling based on $T_{\infty}/T_w$}
\label{sub:temperature_ratio}

So far, $\Delta U_I^+$ in figures \ref{fig:mean}(e,f) and \ref{fig:hama}(a,b) are presented as a function of viscous-scaled $k$ or $k_s$. Here, the viscous lengthscale $U_{\tau}/\nu_w$ is a quantity defined by flow properties at the wall, $U_{\tau} = \sqrt{\tau_w/\rho_w}$. Previous studies suggested that a better collapse to Nikuradse's formula in incompressible flow regime was observed if flow properties at another wall-normal location related to the roughness were used to scale $k$ or $k_s$ instead of those at the wall. In the cube roughness (figure \ref{fig:roughness_type}\textbf{D}) of \cite{modesti2022}, a correction factor based on the kinematic viscosity ratio $\nu_k^+ = \nu_k/\nu_w$ was introduced to the cube physical height, $k_{\ast} = k /\nu_k^+$ (subscript `k' denotes properties at the tip of the roughness elements), such that a collapse to Nikuradse's formula is observed when $\Delta U_I^+$ (for various transformations) was scaled by $k_{s\ast}^+ = 1.9 k U_{\tau}/\nu_k$, with $k_s/k = 1.9$ is the $k_s$ obtained from numerical simulations of the same roughness at $M = 0.3$. A similar scale of $k_{\ast}^+ = (k \sqrt{\tau_w/\rho_k})/\nu_k$ was utilised for the cube and 2D traverse bar roughness (figure \ref{fig:roughness_type}\textbf{B}) of \cite{neeb2016}. 

Unfortunately, these correction factors based on the properties at the tip of the roughness elements are not able to collapse $\Delta U_I^+$ for the present test cases in table \ref{tab:cases}. What seems to suit these cases better is a correction factor based on the freestream, rather than the roughness tip, properties: $\nu_{\infty}^+ = \nu_{\infty}/\nu_w$, which is obtained from $T_{\infty}/T_w$ ratio via equation \eqref{eq:sutherland}. Open symbols in figure \ref{fig:hama}(c) shows $\Delta U_I^+$ of all test cases in table \ref{tab:cases}, with various transformations for compressibility effects (table \ref{tab:transform}) across $M$ and $Re_{\tau}$ sweeps, collapses to Nikuradse's formula (within $\pm 10\%$ error marked by the hatched region in the figure) when presented as a function of $k_{\ast}^+ = k U_{\tau}/\nu_{\infty}$.

We further examine this correction factor for the previous studies listed in table \ref{tab:ref}. Finding the correction factor $\nu_{\infty}^+$ for these studies requires some knowledge regarding the freestream-to-wall temperature ratios $T_{\infty}/T_w$ (or bulk-to-wall temperature ratio $T_b/T_w$) and the viscosity models as a function of the temperature profiles. These quantities were not explicitly reported in the majority of studies in table \ref{tab:ref}. Therefore, the temperature ratios $T_{\infty}/T_w$ are calculated from either the temperature profiles (if reported) or from the isentropic relation (if $T_0$, $M_{\infty}$, and $T_w$ were reported). The dynamic viscosity ratio $\mu_{\infty}^+ = \mu_{\infty}/\mu_w$ is then estimated, typically as an exponential function of the temperature profile, $\mu^+ = (T/T_w)^{\omega}$, where $\omega$ is a constant. If the model used to calculate $\mu^+$ was not reported in the study, Sutherland's model (equation \ref{eq:sutherland}), which is also used in present study, is assumed. For previous studies in the hypersonic flow regime, where the static temperature may reach below 100 K, unless the viscosity model was reported, Keyes's model is assumed \citep{keyes1951,roy2006}. With these assumptions, the correction factors $\nu_{\infty}^+$ can be estimated for all previous studies, with the exception of the sandpaper roughness of \cite{goddard1959}, the 2D sine wave of \cite{tyson2013} and \cite{aghaei-jouybari2023}, and the egg carton roughness (figure \ref{fig:roughness_type}\textbf{E}) of \cite{aghaei-jouybari2023}. Gray-filled symbols in figure \ref{fig:hama}(c) show the reported $\Delta U_I^+$ as a function of $k_{\ast}^+ = k U_{\tau}/\nu_{\infty}$ for test cases in table \ref{tab:ref} (where $\nu_{\infty}^+$ can be estimated). Unlike the present study, the $\nu_{\infty}^+$ correction factor does not collapse the majority of $\Delta U_I^+$ of previous studies to Nikuradse's formula (within $\pm 10\%$ error from the formula, shown by hatched region in figure \ref{fig:hama}c), except for the P80-grit sandpaper of \cite{latin2000}, test cases at $M_b = 2$ of \cite{modesti2022}, and a test case at $M_{\infty} = 2.25$ of \cite{wang2024}. 

Due to poor performance of the correction factor $\nu_{\infty}^+$ to scale the roughness physical height $k$ in the $x$-axis of figure \ref{fig:hama}(c), we develop another correction factor for $\Delta U_I^+$ (the $y$- instead of $x$-axis of figure \ref{fig:hama}c). Considering the relation between the skin friction coefficient $C_f$ and $\Delta U^+$ in equation \eqref{eq:hama_cf}, we introduce a new correction factor for $\Delta U_I^+$, $\sqrt{1/F_c} \Delta U_I^+$, where $F_c$ is the correction factor for smooth-wall TBL skin friction, $F_c C_f$, derived from the VD II transformation \cite{vandriest1956}. For the present experimental data, $F_c$ is calculated using equation \eqref{eq:hopkins} as a function of $T_{\infty}/T_w$. Open symbols in figure \ref{fig:hama}(d) shows the corrected $\Delta U_I^+$ as a function of $k^+$ for all test cases in table \ref{tab:cases}, of various transformations (table \ref{tab:transform}) across $M$ and $Re_{\tau}$ sweeps. All test cases collapse to Nikuradse's formula (within $\pm 10\%$ error marked by the hatched region in the figure) when the correction factor $\sqrt{1/F_c}$ is applied to the experimental data. 

In line with the previously tested correction factors, the $\sqrt{1/F_c}$ correction factor is tested for previous studies in table \ref{tab:ref}. Due to various wall conditions of the experiments and simulations (either adiabatic, isothermal, or a constant cooling rate is applied), a more general form of VD II transformation \citep{vandriest1956} was derived by \cite{hopkins1971} to obtain the correction factor $F_c$ 
\begin{equation}
    F_c = \frac{r \frac{\gamma-1}{2} M_{\infty}^2}{(\arcsin{\alpha} + \arcsin{\beta})^2} 
    \label{eq:hopkins_gen}
\end{equation}
where
\begin{align}
    \alpha &= \frac{2A^2 - B}{\sqrt{4A^2 +B^2}} \quad \mathrm{and} \quad \beta = \frac{B}{\sqrt{4A^2 + B^2}} \\
    A &= \sqrt{\frac{r \frac{\gamma-1}{2} M_{\infty}^2}{T_w/T_{\infty}}} \quad \mathrm{and} \quad B = \frac{1+ r \frac{\gamma-1}{2} M_{\infty}^2 - T_w/T_{\infty}}{T_w/T_{\infty}}  
\end{align}
The temperature ratio $T_{\infty}/T_w$ is obtained either from the temperature profiles or the isentropic relation, as described in the steps to obtain the previous correction factor $\nu_{\infty}^+$. For studies conducted on adiabatic walls (whether assumed or measured directly, as marked by \circled[b] and \circled[c] in table \ref{tab:ref}), equation \eqref{eq:hopkins_gen} reduces to \eqref{eq:hopkins}. Gray-filled symbols in figure \ref{fig:hama}(d) show the $\Delta U_I^+$ corrected by $\sqrt{1/F_c}$ as a function of $k^+$. In general, this correction factor performs better than scaling the $x$-axis by $k_{\ast} = k/\nu_{\infty}^+$, shown in figure \ref{fig:hama}(c), for collapsing $\Delta U_I^+$ onto Nikuradse's formula. Within $\pm 10\%$ error of the formula (hatched region in figure \ref{fig:hama}d), the drag of the following test surfaces can be estimated: all sandpaper roughnesses of \cite{reda1975} and two sandpaper grits (P36 and P20) of \cite{latin2000}, a test case each of the 2D traverse bars of \cite{berg1979}, \cite{latin2000}, and \cite{sahoo2009}, the cube roughnesses of \cite{latin2000}, as well as \cite{modesti2022} for all $M_b = 2$ test cases and only the higher $Re_{\tau}$ cases at $M_b = 4$, the egg carton roughness of \cite{wang2024} at $M_{\infty} \leq 3.5$, and the mesh roughness of \cite{sahoo2009}.           

\section{Discussions}
\label{sub:discuss}

It is not clear at this point why some correction factors that collapse either $\Delta U_I^+$, $k^+$, or $k_s^+$ into Nikuradse's formula perform better or worse than others. As shown in \S\ref{sub:match} and figure \ref{fig:hama}(b), when the roughness type and geometry (i.e. $\delta/k$) are matched to an equivalent study in incompressible flow and the equivalent $k_{si}$ is utilised, there is no clear indicator as to why an estimated $k_{si}$ for a certain roughness can collapse $\Delta U_I^+$ of a rough surface to Nikuradse's formula while another $k_{si}$ for another roughness cannot. Adding flow conditions into consideration, in this case either $Re_{\tau}$ or $\Delta U^+$, does not yield clearer criteria for roughness matching that guarantee the success of the estimated $k_{si}$. This suggests that there are other factors at play, such as compressibility, that have not been considered by this method.        

Some measures of compressibility are taken into account in \S\ref{sub:temperature_ratio}. The first is a correction factor on the $x$-axis of $\Delta U_I^+$ as a function of $k_s^+$ or $k^+$. Typically, this involves some viscosity and/or density correction factor of $k_s^+$ or $k^+$ (where $k_s$ is obtained from an equivalent incompressible test case) such that $\Delta U_I^+$ (obtained from the transformed velocity profile) collapses onto Nikuradse's formula. Here, the density and viscosity profiles are obtained from the temperature profile, with an assumed viscosity model (see for example, equation \ref{eq:sutherland}). This approach has been documented in previous studies, such as \cite{neeb2016} and \cite{modesti2022} for cube and 2D traverse bar roughnesses (figures \ref{fig:roughness_type}\textbf{B} and \textbf{C}), using density and/or kinematic viscosity at the tip of the roughness element, $\rho_k$ and $\nu_k$. To the best of our knowledge, these correction factors were empirical (i.e. not derived analytically) and have not been tested on other roughnesses in other studies. With the present experimental data (table \ref{tab:cases}), scaling $k^+$ with the \emph{freestream}-to-wall air viscosity ratio $\nu_{\infty}^+ = \nu_{\infty}/\nu_w$, rather than the tip of the roughness element, collapses the data onto Nikuradse's formula in the fully rough regime (equation \ref{eq:hama}). The same scaling factor (using the freestream viscosity) has a very limited success when tested on data from previous studies (table \ref{tab:ref}) within transitionally and fully rough regimes, as shown in figure \ref{fig:hama}(c). Similar to the other correction factor, which uses air properties at the roughness tip, the freestream correction factor is also determined empirically.    

A more promising correction factor is one on the $y$-axis, $\Delta U_I^+$ multiplied by $\sqrt{1/F_c}$ (figure \ref{fig:hama}d), where $F_c$ a function of freestream-to-wall temperature ratio, $T_{\infty}/T_w$ (equations \ref{eq:hopkins} and \ref{eq:hopkins_gen}). $F_c$ is a correction factor derived from the \cite{vandriest1956} II transformation for $C_f$ of \emph{smooth}-wall boundary layers. Using this correction factor, the present experimental data collapse onto Nikuradse's formula in the fully rough regime (table \ref{tab:cases}). When tested for the data from previous studies (table \ref{tab:ref}), the $\sqrt{1/F_c}$ has better success in collapsing the $\Delta U_I^+$ compared to the other correction factors tested in the study (figure \ref{fig:hama}d): using the incompressible equivalent $k_s$ (\S\ref{sub:match}) and scaling $k^+$ (or $k_s^+$) by the freestream viscosity (\S\ref{sub:temperature_ratio}). Notably, these studies outlined in table \ref{tab:ref} span supersonic to hypersonic freestream (or bulk) velocities, Reynolds numbers, transitionally to fully rough regimes, and perhaps more crucially, various techniques to obtain the wall-shear stresses (whether through direct measurements or fitting methods, as is the case with present experimental data), as well as various wall conditions (direct $T_w$ measurements, assumed $T_w$, or wall treatments, e.g. cooled with a constant cooling rate). However, similar to the previous correction factors for $k^+$ or $k_s^+$ involving some density or viscosity of the flow, either at the tip of the roughness or in the freestream, this correction factor is, too, empirical. In fact, the Hama roughness function $\Delta U_I^+$ is obtained by vertically shifting a transformed velocity profile (table \ref{tab:transform} and \S\ref{sub:wss}), and thus multiplying this quantity by $\sqrt{1/F_c}$ is, in a sense, performing the transformation twice. 

\begin{figure}
    \centering
    \includegraphics[width=13.5cm, keepaspectratio]{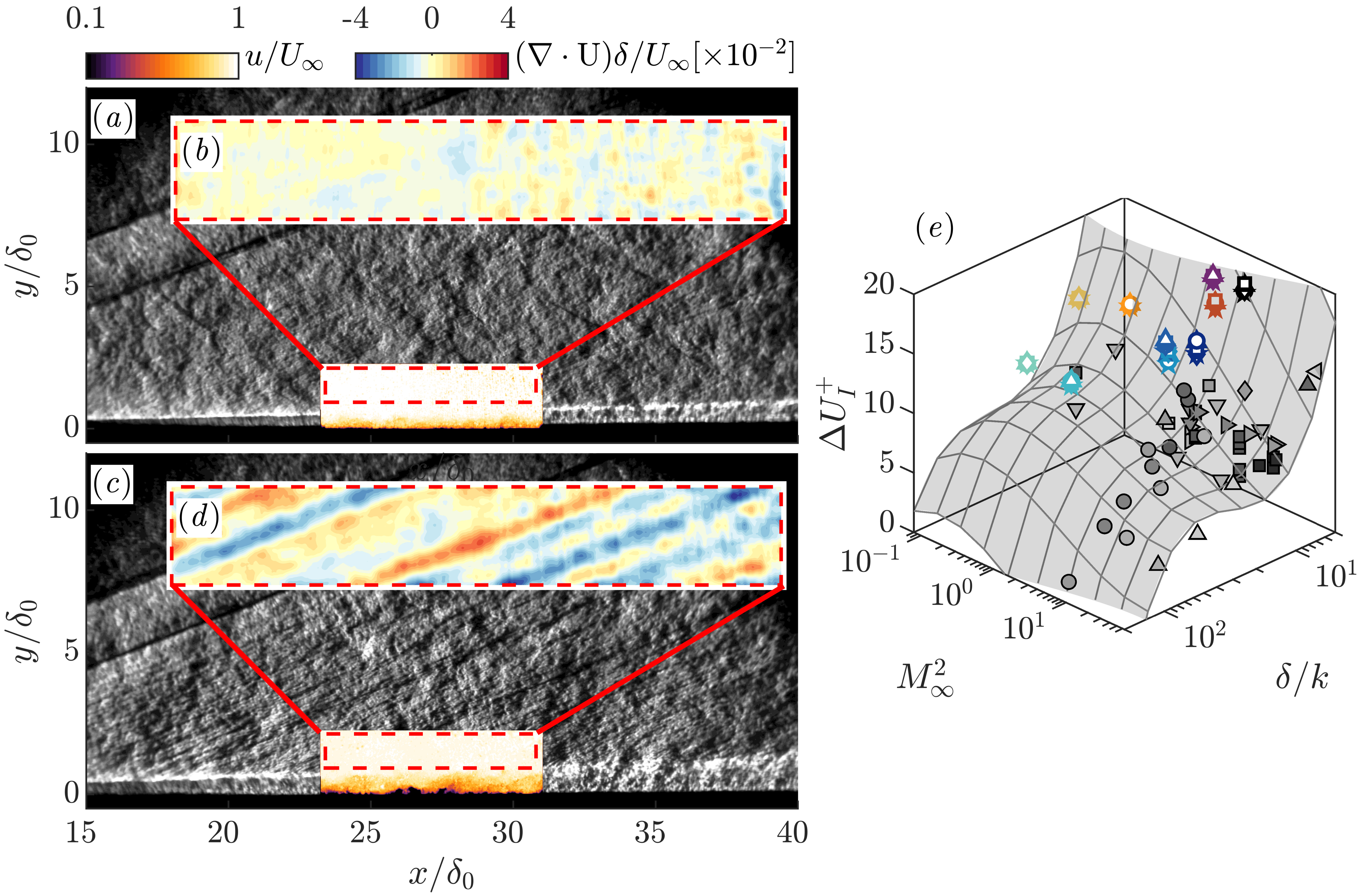}
    \caption{Instantaneous Schlieren image of test case (a) SW and (c) P24 at $M = 2.9$, overlaid by an instantaneous streamwise velocity $u/U_{\infty}$ obtained from PIV measurements (not simultaneously acquired), where $\delta_0$ is the reference boundary-layer thickness (case SW at $M = 0.3$, see also figure \ref{fig:setup}c). \emph{Inset:} time-averaged velocity divergence $\nabla \cdot \mathbf{U}$ normalised by $\delta/U_{\infty}$ of the same (b) SW and (d) P24 test cases at the freestream, the area marked by red dashed lines in (a, c). (e) Surface of $\Delta U_I^+$ as a function of $\log M_{\infty}^2$ (or $M_b^2$ for channel flow cases) and $\log(\delta/k)$, obtained from fitting the present experimental results (open coloured symbols in table \ref{tab:cases}) and data from previous studies (gray-filled symbols in table \ref{tab:ref}).}
    \label{fig:discuss}
\end{figure}


Perhaps, the flow fields should be re-examined. Figure \ref{fig:discuss}(a) shows an instantaneous Schlieren image in the $x$--$y$ plane of the SW test case at $M = 2.9$, taken by a pco.Dimax S4 high-speed camera and overlaid by an instantaneous $u/U_{\infty}$ flow field obtained from the 2D2C PIV measurements at the same plane (not simultaneously acquired). The leading-edge oblique shockwave is captured by the Schlieren image, visible on the left-hand side of figure \ref{fig:discuss}(a). Inset (b) of the same figure captures the time-averaged velocity divergence $\nabla \cdot \mathbf{U} = \partial U/\partial x + \partial V/\partial y$ at the freestream (red dashed lines in figure \ref{fig:discuss}a), obtained from PIV measurements and normalised by $\delta/U_{\infty}$. For the SW test case, the divergence at the freestream approaches zero, as shown in figure \ref{fig:discuss}(b). By contrast, the instantaneous Schlieren image of the P24 test case at the same nominal $M = 2.9$ (figure \ref{fig:discuss}c) reveals a succession of shockwaves (or more accurately, bow shocks) in front of the roughness elements in the streamwise direction, which is related to the additional wave drag in compressible flows. Furthermore, these waves are persistent and extend well into the freestream: they are also observed in the time-averaged flow field, as shown by the mean velocity divergence in the freestream (figure \ref{fig:discuss}d). The current framework for drag characterisation in compressible, rough-wall TBLs (figure \ref{fig:framework}) does not account for the additional wave drag inferred from figure \ref{fig:discuss}(c,d), as the velocity profiles are stretched with transformations developed or derived from \emph{smooth}-wall, rather than \emph{rough}-wall, TBLs (examples of such transformations are outlined in table \ref{tab:transform}). In the future, a custom transformation for a rough-wall TBL, which accounts for the properties of a particular roughness, should be considered. 

Ultimately, drag characterisation in compressible, rough-wall TBLs aims to build a ``map" of a drag quantity (either $C_f$ or $\Delta U_I^+$, as shown in figure \ref{fig:discuss}e) as a function of roughness properties (or a custom compressibility transformation for a rough wall), inflow Mach number $M_{\infty}$ (or $M_b$ for channel flow cases), and possibly other factors (e.g. wall conditions). Data from previous studies, both experimental and computational (listed in table \ref{tab:ref}), are unfortunately limited. And while the present experimental data bridge the data gap in the high $Re$ regime, with independent sweeps of $M$ and $Re$ spanning the subsonic, transonic, and supersonic flow regimes, direct measurements of wall-shear stress, wall temperatures, and temperature profiles are not carried out due to limitations of the experimental facility. These measurements should be carried out in future studies to ensure an accurate drag characterisation. However, measuring properties at the wall, or, as suggested by some studies, at the tip of the roughness, is non-trivial. This is especially true for non-homogeneous, irregularly-shaped roughness, akin to the sandpaper roughness tested in this study, which is more likely to arise from impact and ablation during the operation of a high-speed vehicle. 

\section{Conclusions and future work}

Experiments are carried out for TBLs developing over two types of sandpaper rough walls (P60- and P24-grit) in order to establish a framework for drag characterisation in compressible TBLs using the existing framework \citep{nikuradse1933,hama1954} and information for the same rough surfaces in incompressible TBLs (figure \ref{fig:framework}). The test cases span both incompressible and compressible flow regimes ($0.3 \leq M \leq 2.9$) and $7427 \leq Re_{\tau} \leq 30292$, with an independent variation of $Re_{\tau}$ at a constant $M = 2$. Due to the limitations of the test facility, the wall-shear stress is estimated using a fitting method outlined in \S\ref{sub:wss} and an adiabatic wall is assumed (\S\ref{sub:wss}).

Present results suggest that all test cases in table \ref{tab:cases} fall into the fully rough regime and that the Hama roughness function (or the momentum deficit) $\Delta U_I^+$ in the compressible flow regime is relatively independent of the velocity transformations to account for the compressibility effect (table \ref{tab:transform}, figure \ref{fig:hama}a). The log-law intercept of $B-B_{FR}$ (equation \ref{eq:hama}), however, seems to increase proportionally with $M$. This is not taken into account in the existing log relation in the fully rough regime (equation \ref{eq:hama}) for incompressible flow. Therefore, to be able to utilise the existing framework for drag characterisation, either the roughness lengthscale $k^+$ or $k_s^+$ or the $\Delta U_I^+$ must be scaled by a factor.

Three scaling factors are introduced and tested for the present experimental data and the available data from previous studies described in table \ref{tab:ref} involving various roughness types illustrated in figure \ref{fig:roughness_type}. The first is an equivalent $k_s$ obtained from a similar incompressible flow study (the same roughness type and comparable roughness length-scale $\delta/k$), followed by a correction factor in $k$, $k_{\ast} = k/\nu_{\infty}^+$, and finally a correction factor for $C_f$ in the momentum deficit $\Delta U_I^+$, $\sqrt{1/F_c} \Delta U_I^+$, where $F_c$ is a function of $T_{\infty}/T_w$ (equations \ref{eq:hopkins} and \ref{eq:hopkins_gen}). When tested for the current and previous compressible flow data, the first method has limited success (figure \ref{fig:hama}b). Both second and third methods collapse the present experimental data into the log relation (figures \ref{fig:hama}c,d). The latter shows more promising results when tested for data from previous studies (figure \ref{fig:hama}d).

However, it should be noted that all correction factors tested in this study are obtained empirically and that the drag and wall temperature are not directly measured. Furthermore, the current drag characterisation framework (Figure \ref{fig:framework}) uses compressibility transformations derived for smooth-wall TBLs (table \ref{tab:transform}) and the proposed correction factor $F_c$, which shows the most promising result, was originally derived for the correction of the smooth-wall $C_f$. In the future, direct wall-shear stress and wall temperature measurements should be conducted to ensure an accurate drag characterisation. Using these data, a custom transformation for a rough-wall TBL should be developed, which accounts for the properties of a particular roughness, the wave drag formed by the shock generated by the roughness elements, and the wall conditions.     


\backsection[Funding]{The study is funded by the German Research Foundation (DFG, grant no. 448354709). DDW acknowledges funding from the Leverhulme Early Career Fellowship (grant no. ECF-2022-295). BG acknowledges funding from EPSRC (Grant Ref: EP/W026090/1).}

\backsection[Declaration of interests]{The authors report no conflict of interest.}

\backsection[Data availability statement]{The data that support the findings of this study will be made available upon publication at the University of Southampton's repository.}

\backsection[Author ORCIDs]{\\D. D. Wangsawijaya, https://orcid.org/0000-0002-7072-4245; \\R. Baidya, https://orcid.org/0000-0002-6148-602X; \\S. Scharnowski, https://orcid.org/0000-0002-6452-2954; \\B. Ganapathisubramani, https://orcid.org/0000-0001-9817-0486; \\C. J. K{\"ahler}, https://orcid.org/0000-0001-9336-2091.}


\bibliographystyle{jfm}
\bibliography{biblio}


\end{document}